\documentclass[%
 reprint,
 superscriptaddress,
preprintnumbers,
nofootinbib,
 amsmath,amssymb,
 aps,
]{revtex4-1}

\usepackage{amsmath}
\usepackage{amssymb}
\usepackage{amsthm}
\usepackage{MnSymbol}
\usepackage{nicefrac}
\usepackage{amsfonts}
\usepackage{dsfont}

\usepackage[markup=nocolor]{changes}
\usepackage{natbib}

\usepackage{graphicx}
\graphicspath{{../Figures/}}  

\usepackage{dcolumn}
\usepackage{bm}

\usepackage{color}
\usepackage{slashed}
\definecolor{mygreen}{rgb}{0, 0.6, 0}
\usepackage{empheq}

\newcommand{\bea}{\begin{eqnarray}}
\newcommand{\eea}{\end{eqnarray}}
\newcommand{\be}{\begin{equation}}
\newcommand{\ee}{\end{equation}}

\def\s0#1#2{\mbox{\small{$ \0{#1}{#2} $}}}
\def\0#1#2{\frac{#1}{#2}}

\def\eq#1{(\ref{#1})}

\begin{document}

\title{Quantum-gravity effects on a Higgs-Yukawa model}

\author{Astrid Eichhorn} \affiliation{Blackett Laboratory, Imperial
  College, London SW7 2AZ, United Kingdom}
  \preprint{Imperial/TP/2016/AE/01}

\author{Aaron Held}
\affiliation{Institut f\"ur Theoretische
  Physik, Universit\"at Heidelberg, Philosophenweg 16, 69120
  Heidelberg, Germany} 

\author{Jan M. Pawlowski}
\affiliation{Institut f\"ur Theoretische
  Physik, Universit\"at Heidelberg, Philosophenweg 16, 69120
  Heidelberg, Germany} 
\affiliation{ExtreMe Matter Institute EMMI, GSI, Planckstr.~1,
  D-64291 Darmstadt, Germany}

\begin{abstract}
  A phenomenologically viable theory of quantum gravity must
  accommodate all observed matter degrees of freedom and their
  properties. Here, we explore whether a toy model of the Higgs-Yukawa
  sector of the Standard Model is compatible with asymptotically safe
  quantum gravity. We discuss the phenomenological implications of our
  result in the context of the Standard Model.

  We analyze the quantum scaling dimension of the system, and find an
  irrelevant Yukawa coupling at a joint gravity-matter fixed
  point. Further, we explore the impact of gravity-induced couplings
  between scalars and fermions, which are non-vanishing in
  asymptotically safe gravity.
 
\end{abstract}

\pacs{Valid PACS appear here}

\maketitle


\section{Introduction}
\label{sec:intro}

Asymptotic safety \cite{Weinberg:1980gg} provides a framework in which
a local quantum field theory of the metric - or related gravitational
degrees of freedom, such as, e.g., the vielbein - could be
established. Compelling hints for the viability of the
asymptotic-safety scenario have been uncovered in pure gravity
\cite{Reuter:1996cp,
  Dou:1997fg,Souma:1999at,Reuter:2001ag,Lauscher:2001ya,Lauscher:2002sq,%
Litim:2003vp,Pawlowski:2003sk,Fischer:2006fz,Machado:2007ea,Eichhorn:2009ah,
  Codello:2006in,Codello:2008vh,Benedetti:2009rx,Manrique:2009uh,Eichhorn:2010tb,
  Groh:2010ta,Manrique:2010mq,Manrique:2010am,Daum:2010qt,Manrique:2011jc,%
Donkin:2012ud,Christiansen:2012rx,Rechenberger:2012dt,Benedetti:2012dx,%
Dietz:2012ic,Codello:2013fpa,Falls:2013bv,Benedetti:2013jk,Daum:2013fu,Ohta:2013uca,%
Demmel:2014sga,Falls:2014tra,Becker:2014qya,Christiansen:2014raa,Falls:2015qga,%
Falls:2015cta,Dietz:2015owa,Gies:2015tca,Demmel:2015oqa,Christiansen:2015rva,Gies:2016con,Labus:2016lkh}, 
for reviews see \cite{Niedermaier:2006wt,Niedermaier:2006ns,Percacci:2007sz,Litim:2008tt,Litim:2011cp,Reuter:2012id,Reuter:2012xf}; similar ideas have been explored in other quantum field theories, see, e.g., \cite{Gies:2009hq,Braun:2010tt,Litim:2014uca}.
These results suggest that quantum fluctuations of gravity could
induce an interacting Renormalization Group (RG) fixed point,
generalizing the success-story of asymptotic freedom to a quantum
gravitational setting.  Yet, a fundamental model of spacetime that is
viable in our universe must be compatible with the existence and
properties of the observed matter degrees of freedom.  First results
suggest that asymptotic safety in gravity persists if the matter
fields of the Standard Model and some of its extensions are coupled
minimally to gravity
\cite{Dona:2013qba,Dona:2014pla,Meibohm:2015twa,Dona:2015tnf}. Here,
we focus on the converse question, namely the effect of quantum
gravity on matter fields and their interactions in the ultraviolet
(UV). The Standard Model is a low-energy effective field theory, which
presumably features perturbative Landau poles and a triviality problem
beyond perturbation theory in the Higgs sector and the U(1)
hypercharge sector.  Within the asymptotic-safety scenario local
quantum field theory remains a valid framework up to arbitrarily high
momentum scales, requiring a quantum-gravity induced Renormalization
Group fixed point for matter.  First hints for its existence have been
found in \cite{Zanusso:2009bs,Narain:2009fy,Daum:2009dn,Vacca:2010mj,%
  Daum:2010bc,Harst:2011zx,Folkerts:2011jz,Eichhorn:2011pc,%
  Eichhorn:2012va,Oda:2015sma,Meibohm:2016mkp}.

In this work, we focus on a toy model of the Yukawa sector of the
Standard Model, a matter system with a massless Dirac fermion and a
real scalar interacting via a Yukawa term with coupling $y$, in the
presence of quantum gravity. Within canonical power counting, the
Yukawa coupling $y$ corresponds to a marginally irrelevant
coupling. If the coupled matter-gravity system is asymptotically safe
and the Yukawa coupling remains irrelevant, this results in a
prediction for the low-energy limit: The value of the latter is
dominated by the UV-relevant couplings in the vicinity of the UV-fixed
point, which determine the RG-trajectory from the UV-regime to the low
energy regime. In turn, UV-irrelevant couplings correspond to a
UV-repulsive direction of a fixed point, and their value governs the
approach to the critical surface at low energies. Thus, in this
scenario most values of $y$ in the infrared (IR) are incompatible with
asymptotic safety. A relevant Yukawa coupling most
  likely implies that its observed low-energy value can be
  accommodated within an asymptotically safe UV completion. On the
  other hand, an irrelevant Yukawa coupling makes the model more
  predictive, and can only accommodate one particular value of the
  Yukawa coupling. If this does not match the observed value, this
  particular UV completion is ruled out. The arguably most interesting
  case is realized if the Yukawa coupling is irrelevant, and the
  predicted low-energy value agrees with observations. This scenario
  would showcase how an interacting fixed point can have an enhanced
  predictive power over a perturbative setting, where the Yukawa
  couplings are free parameters.

 In this spirit
it has been conjectured that the
Higgs mass could be predicted within asymptotic safety
\cite{Shaposhnikov:2009pv, Bezrukov:2012sa}.  Two conditions need to
be satisfied for this scenario to work in the Standard Model: Firstly,
the Higgs self-interaction must become irrelevant at the fixed
point. Secondly, the top-Yukawa coupling needs to be relevant, if the
corresponding fixed point lies at a vanishing value of the
coupling. Otherwise, the UV-repulsive fixed point at $y=0$ is
difficult to reconcile with the value of the top-Yukawa coupling at
the Planck scale, which is significantly larger than zero in the
Standard Model \cite{Buttazzo:2013uya}.

The main goals of this paper are the following two: Firstly, we study the
quantum gravitational correction to the critical exponent of the
Yukawa coupling at its fixed point to determine whether the low-energy
value of the Yukawa coupling could be predicted in a matter-gravity model
 in Section~\ref{sec:gauge}. Secondly, we take a more
detailed look at the properties of a joint matter-gravity fixed point.
In particular, we focus on quantum-gravity induced interactions, and
the shift of a possible matter fixed point to an asymptotically safe
instead of an asymptotically free one in Section~\ref{sec:quartic}.


\section{Yukawa theory coupled to quantum gravity}
\label{sec:frg}
\subsection{Effective action}
\label{sec:eff_action}
We will analyze the momentum-scale running of the effective action
$\Gamma_k$ of a Yukawa theory coupled to gravity in the presence of an
infrared cutoff scale $k$. More specifically, we will monitor the
scale-dependence of matter couplings in the vicinity of the
asymptotically safe UV fixed point of the theory. The theory consists
of a Dirac fermion $\psi$ and a real scalar $\phi$ coupled to a
fluctuating metric. The flowing action $\Gamma_k$ of the model is
parameterized as
\begin{align}\nonumber 
  \Gamma_k[\bar g,\Phi] =&\, \frac{Z_{\phi}}{2} \int d^4x \sqrt{g}
  \left(g^{\mu \nu} \partial_{\mu}\phi \partial_{\nu}\phi
    + m_\phi^2 \phi^2\right) \\[2ex]
  &\,+ i Z_{\psi}\int d^4 x \sqrt{g} \bar{\psi}\slashed{\nabla} \psi \nonumber\\[2ex]
  &\,+iZ_{\psi}Z_{\phi}^{1/2}y\,\int d^4x \sqrt{g}\, \phi\,\bar\psi\psi \nonumber \\[2ex]
  &\,+\Gamma_{k,\rm ho}[\bar g,\Phi] + \Gamma_{k,\rm grav}[\bar
  g,\Phi]\,,
\label{eq:truncation}
  \end{align}
  with scale-dependent couplings $y(k)$ and $m_{\phi}^2(k)$ and
  wave-function renormalizations $Z_{\phi}(k)$, $Z_{\psi}(k)$. The
  Dirac term contains the spin connection, see
  App.~\ref{app:matter-vertices}. The effective action depends on an
  auxiliary background metric $\bar g_{\mu\nu}$ and the
  fluctuating quantum fields $\Phi$ with
\begin{align}
  \Phi = (h_{\mu \nu},c_{\nu},\bar c_{\nu},\psi,\bar \psi,
  \phi)\,.\label{eq:Phi}
\end{align}
In \eq{eq:Phi} the field $h$ carries the metric fluctuations and the
corresponding pure-gravity dynamics is contained in $\Gamma_{k,\rm
  grav}$. The full metric $g_{\mu\nu}=g_{\mu\nu}(\bar g,h)$ in the
first three lines of \eq{eq:truncation} is chosen to be
\begin{align} \label{eq:gbargh}
  g_{\mu \nu}=Z^{1/2}_{\bar g}\, \bar{g}_{\mu \nu} +\sqrt{G}\,
  Z_h^{1/2}\, h_{\mu \nu}\,, 
\end{align}
with Newton coupling $G$. The wave function renormalizations $Z_{\bar
  g}$ and $Z_h$ carry the cutoff-scale dependence of the respective
kinetic terms similar to those of the scalar and fermion. The
parameterization \eq{eq:truncation} with \eq{eq:gbargh} leaves us with
RG-invariant but scale-dependent couplings $y$ and $G$, while the
RG-scaling of the vertices is  carried  entirely by the corresponding
$Z$-factors. The factor $\sqrt{G}$ leaves the fluctuation field $h$
with the canonical dimension of a bosonic field such that powers of
$G$ occur in all matter-gravity and pure-gravity vertices.  

The presence of the background metric is required for two reasons:
Firstly, the use of an RG-approach in gravity requires to set a scale
in order to distinguish the ``high-momentum modes" of the model. This
is possible with the help of the background-covariant
Laplacian. Secondly, our approach requires the specification of a
propagator for the metric, which, as in any gauge theory, requires the
introduction of a gauge-fixing term. Here, we will use a family of
gauge fixing functionals with respect to the background metric.

In the present setting classical diffemorphism invariance is encoded
in Slavnov-Taylor identities. They imply that the quantum effective
action cannot be expanded in diffeomorphism-invariant terms. This
entails that the first three lines in \eq{eq:truncation} are
accompanied by higher order terms $\Gamma_{k,\rm ho}$ dictated by the Slavnov-Taylor
identities. These terms are neglected in the present work. A more
detailed discussion of 
background independence is given in App.~\ref{app:diffeo}.

The propagator of metric fluctuations is obtained from the 
Einstein-Hilbert action with general covariant gauges and the
  same linear metric split, cf.~\eqref{eq:gbargh}.  The classical
action reads
\begin{equation}\label{eq:EHaction}
  S_\text{EH}=\frac{1}{16\pi G}\int\mathrm d^4x
  \sqrt{g}\,(2\Lambda-R)+S_\text{gf}+S_\text{gh}\,.
\end{equation}
and we employ the linear gauge fixing condition 
\begin{align}
\label{eq:gauge-condition}
F_\mu = \left( \delta_\mu^\rho\bar D^\sigma - \frac{1+\beta}{
    4}\bar{g}^{\rho\sigma} \bar D_\mu \right) h_{\rho\sigma}\,,
\end{align}
where $\bar D$ is the covariant derivative w.r.t.\ the background
metric $\bar g_{\mu\nu}$. This results in a gauge-fixing term, 
\begin{align}
  S_\text{gf} = \frac{1}{32\pi G \alpha}\int d^4x \sqrt{\bar
    g}\bar{g}^{\mu\nu}F_\mu F_\nu, 
 \end{align}
 accompanied by the standard Faddeev-Popov-ghost term. We thus have a
 two-parameter family of gauges labelled by $(\alpha,\beta)$.  Given
 an expansion in powers of the fluctuating graviton $h$, we obtain the
 quadratic part of the pure-gravity effective action from which we
 construct the scale-dependent full metric propagator, for details see
 App.~\ref{app:gravity}. In the present work we will not evaluate the
 gravitational RG flow explicitly, but instead utilize results for
 pure quantum gravity, and quantum gravity coupled to free
 matter. Thus gravitational couplings such as $G$ and the
 graviton-mass parameter $\mu_h$ (defined in \eqref{eq:hh-app}) constitute free parameters of our
 equations.

 We close this Section with a discussion of the global symmetries of
 the model in \eq{eq:truncation}. The kinetic terms of the
 scalar and fermion feature a separate $\mathbb{Z}_2$ symmetry under
 which $\phi \rightarrow - \phi$, and a chiral symmetry in the fermion
 sector under which $\psi \rightarrow e^{i \gamma_5 \vartheta} \psi$ and
 $\bar{\psi} \rightarrow \bar{\psi}e^{i \gamma_5 {\vartheta}}$.  The
 Yukawa coupling reduces these separate symmetries to a combined
 discrete chiral symmetry under which the scalar and the fermions
 transform simultaneously and $\vartheta = \pi/2$.  It has the same
 effect a global chiral symmetry would have in the Standard Model,
 forbidding a fermionic mass term.  Further, it restricts the type of
 self-interactions that can be induced by quantum gravity.  All
 induced interactions respect the global symmetries of the matter
 sector; in particular the \emph{separate} symmetries of the fermionic
 and scalar kinetic terms, as these provide the vertices that source
 the induced interactions.  Note that it would be interesting to
 understand whether the general argument, that black-hole
 configurations in the quantum gravitational path integral lead to the
 breaking of global symmetries \cite{Kallosh:1995hi} also applies in
 asymptotic safety and if the preservation of global symmetries that
 we observe is an artifact of the truncation or our choice of
 background and/or signature.

\subsection{Functional renormalization group approach}
\label{sec:FRG}

To explore the scale-dependence of our matter-gravity theory we
employ the non-perturbative functional Renormalization Group (FRG)
approach, for reviews see \cite{Berges:2000ew,Aoki:2000wm,Bagnuls:2000ae,Polonyi:2001se,
Pawlowski:2005xe, Gies:2006wv,Delamotte:2007pf,Braun:2011pp,Rosten:2010vm}.

In this approach the theory is regularized in the infrared below a
momentum scale $k$. This is achieved with an infrared regulator term
in the classical action that underlies the generating functional,
\begin{align}\label{eq:reg}
S_{\rm cl} \to S_{\rm cl} +\012 \int_x \sqrt{\bar g} \, \Phi_I R_{k,IJ}\Phi_J\,, 
\end{align}
where $I,J$ comprise internal indices and species of fields. For
example, for the metric fluctuation $\Phi_1= h$ we have $I=1, \mu
\nu$. The regulator term in \eqref{eq:reg} has to be quadratic in the
fluctuation field in order for the FRG flow equation to be of a
particularly simple one-loop structure. Hence, the regulator contains a
background field dependence with $\sqrt{\bar g}$.  Moreover, it is
chosen such that it suppresses low-momentum fluctuations in the
path-integral, while high-momentum ones are integrated out. In gravity
this again necessitates the specification of a metric, and hence also
$R_k$ depends on the background metric, e.g., via the covariant
background Laplacian $\Delta_{\bar g}$. The explicit form of the
regulators $R_{k,IJ}$ used in the present work is given in
App.~\ref{app:reg}.

Within this framework $\Gamma_k$ interpolates between the microscopic
action for $k\rightarrow\infty$ and the full quantum effective action
for $k\rightarrow 0$. The effective action $\Gamma_k[\bar{g}_{\mu
  \nu},\Phi]$ obeys a one-loop flow equation, the Wetterich equation
\cite{Wetterich:1992yh},
\be
\partial_t \Gamma_k := k \partial_k \Gamma_k = \frac{1}{2} {\rm Tr}\,
\frac{1}{\Gamma_k^{(2)}[\bar{g}_{\mu \nu}, \Phi]+ R_k } \partial_t
R_k.\label{eq:flow} 
\ee 
see also \cite{Ellwanger:1993mw,Morris:1993qb}. In \eq{eq:flow}, the
trace $\rm Tr$ includes a sum over internal indices and species of
fields, including a negative sign for fermions, as well as a
momentum-integration.  $\Gamma_k^{(2)}=\delta^2\Gamma_k/\delta\Phi^2$
denotes the second functional derivatives of the effective action with
respect to $\Phi$. This flow equation can be depicted as a 1-loop
diagrammatic equation, as $1/(\Gamma_k^{(2)}+ R_k)$ corresponds to the
full, field dependent propagator of the theory.  The corresponding
diagrams are not perturbative diagrams but rather fully
non-perturbative depictions of derivatives of the above flow equation
\eqref{eq:flow}.

Quantum fluctuations generate all interactions which are compatible
with the symmetries of the model. For practical purposes, it is
necessary to truncate the space of all action functionals to a
(typically) finite dimensional space. As our truncation, we choose
\eqref{eq:truncation}. In the first part of this work, see
Sec.~\ref{sec:gauge}, we consider a truncation containing the first
three lines in \eqref{eq:truncation} and neglect all additional
terms. 
In a canonical
counting, all scalar-fermion interaction terms beyond this truncation
are irrelevant. While residual interactions at an asymptotically safe
fixed point alter the scaling properties of operators, the canonical
dimensionality can still remain a useful ordering principle. In
particular, if quantum fluctuations shift the critical exponents by
anomalous contributions of $\mathcal{O}(1)$, then only a small number
of couplings can be relevant. This reasoning has been demonstrated to
hold in pure gravity within a truncation based on higher powers of the
scalar curvature, \cite{Falls:2013bv}. Here, we adopt this principle
to motivate our truncation in the matter sector. More formally, our
truncation can be understood as the leading order in a combined
vertex- and derivative expansion, cf.~App.~\ref{app:expansion}.

In the second part, cf.~Sec.~\ref{sec:quartic}, we ask which of the
higher-order couplings contained in $\Gamma_{k,\rm ho}[\bar g,\Phi]$
in \eqref{eq:truncation} are induced by gravity. Specifically,
these are the leading-order terms in a canonical counting, for which
the only possible fixed point \emph{must} be interacting.  The
vertices in the diagrams underlying their beta functions arise from
the \emph{kinetic} terms for scalars and fermions, and must respect
the separate $\mathbb{Z}_2$ symmetry in the scalar and chiral symmetry
in the fermion kinetic terms (cf.~last paragraph in
Sec.~\ref{sec:eff_action}).  The canonically most relevant induced structures
are quartic in the fields, i.e., two-fermion--two-scalar interactions, of order $p^3$. 
A complete basis of the induced operators reads
\begin{align}\nonumber 
&  \Gamma_{k\, \rm ho}[\bar{g}, \Phi]= \Gamma_{k\, \rm induced} \\[2ex] 
= &\, i Z_{\phi}
  Z_{\psi}\, \bar{\mathcal{X}}_{1-}\!\int_x\sqrt{g}\;
  \left[
    \left(\bar\psi\gamma^\mu\nabla_\nu\psi - (\nabla_\nu\bar\psi)\gamma^\mu\psi\right)
	\partial_\mu\phi\partial^\nu\phi \right]
	\notag\\[2ex] 
    + &\, i Z_{\phi}
  Z_{\psi}\,\bar{\mathcal{X}}_{2-}\!\int_x\sqrt{g}\;
  \left[
    \left(\bar\psi\gamma^\mu\nabla_\mu\psi - (\nabla_\mu\bar\psi)\gamma^\mu\psi\right)
	\partial_\nu\phi\partial^\nu\phi \right] \,.
\label{eq:YukawaX}\end{align}
In App.~\ref{app:expansion} we give a more detailed discussion of our
expansion in the matter and gravity sector. This also helps to
understand the technical details as well as the systematics underlying
the derivation of the $\beta$-functions of the matter couplings that
are used for our analysis in Sec.~\ref{sec:gauge} and
Sec.~\ref{sec:quartic}.

\section{Fixed points for the Yukawa coupling}
\label{sec:gauge}
In this section we analyze the UV-stability of the present
Yukawa-gravity system, which serves as a toy model for the top-Yukawa
sector of the Standard Model. It is a first step towards answering the
question, whether asymptotically safe quantum gravity is compatible
with a sizeable top-Yukawa coupling at the Planck scale $M_{\rm Pl}$.
This is required within the Standard Model without additional degrees
of freedom up to $M_{\rm Pl}$. Then, a perturbative evaluation is
viable up to that scale, and yields $y \approx 0.4 / \sqrt{2}$
\cite{Buttazzo:2013uya}.

This is compatible with asymptotic safety within two physically
distinct scenarios. In the first one, the Yukawa coupling features a
UV-attractive fixed point, assuming that its value at the Planck scale
lies within the basin of attraction of the fixed point. In the second
scenario the Yukawa coupling becomes irrelevant at the fixed
point. Then, its low-energy value, i.e., at momentum scales at or
below the Planck scale, is a prediction of asymptotic
safety. Disregarding the curvature of the critical surface, the
fixed-point value would have to correspond to the value at the Planck
scale.

\subsection{Results}\label{sec:results}

\begin{figure}[t]
  \includegraphics[width=0.2\linewidth]{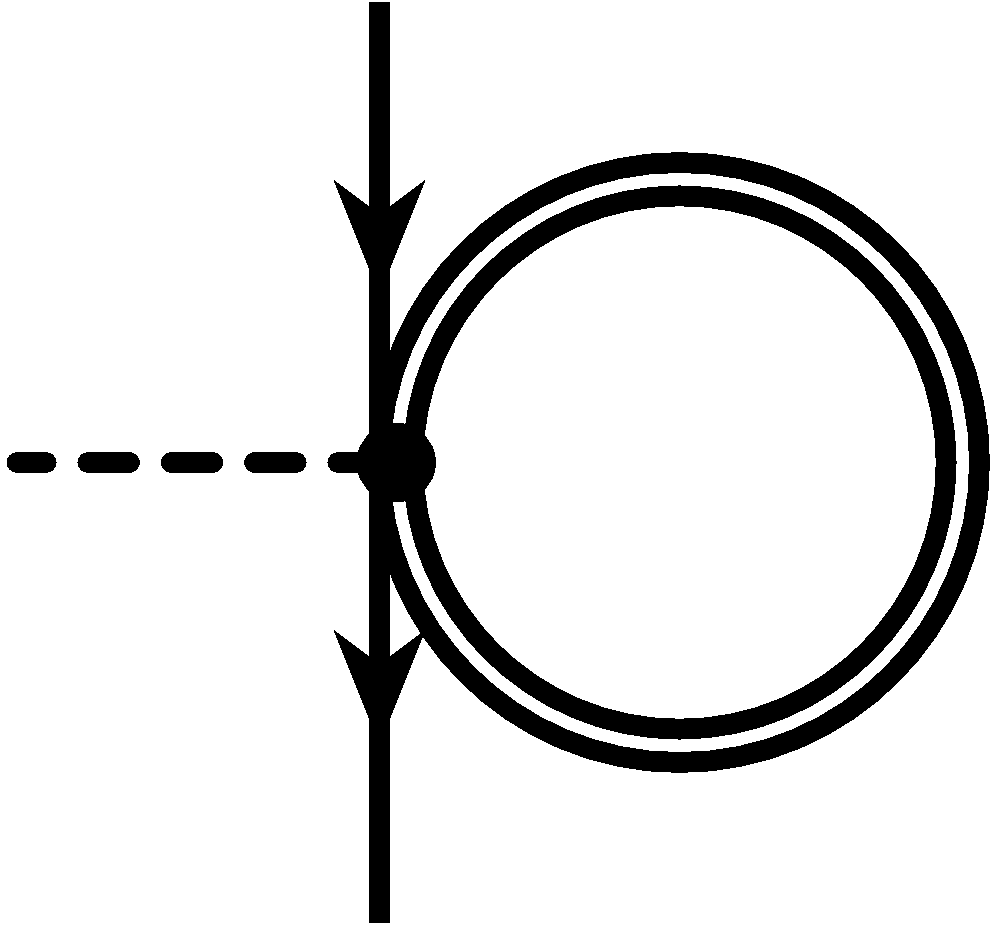}
  \hspace{0.025\linewidth} \includegraphics[width=0.25
\linewidth]{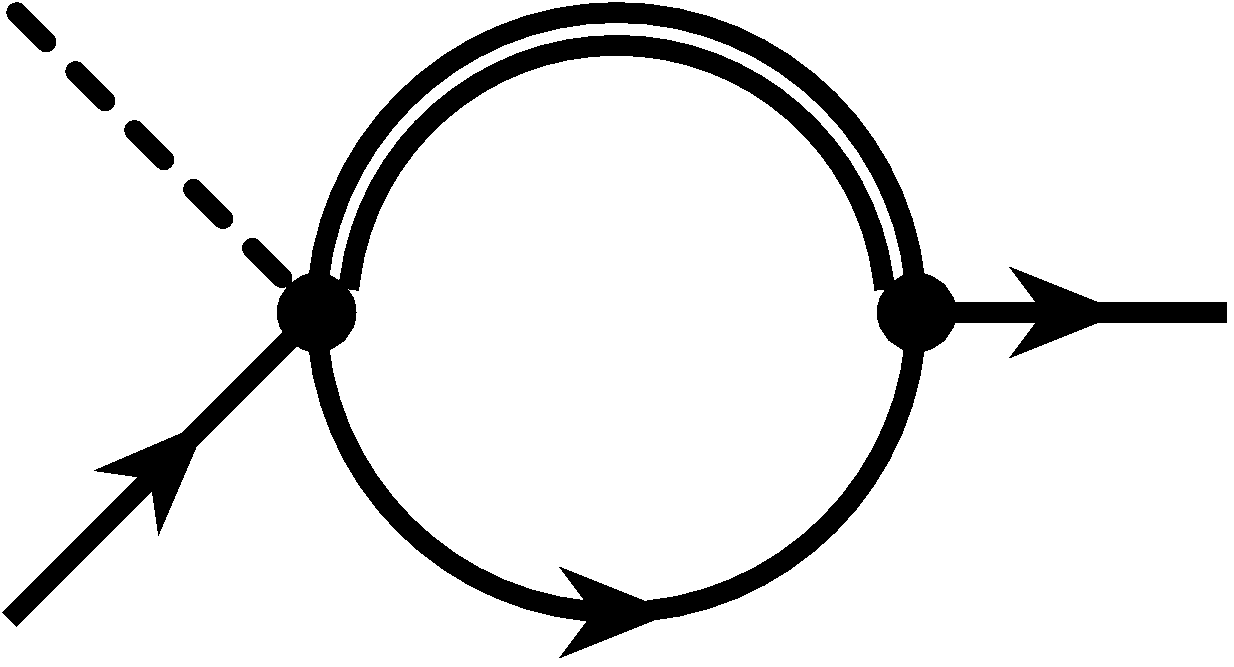}\hspace{0.02\linewidth}
\includegraphics[width=0.25\linewidth]{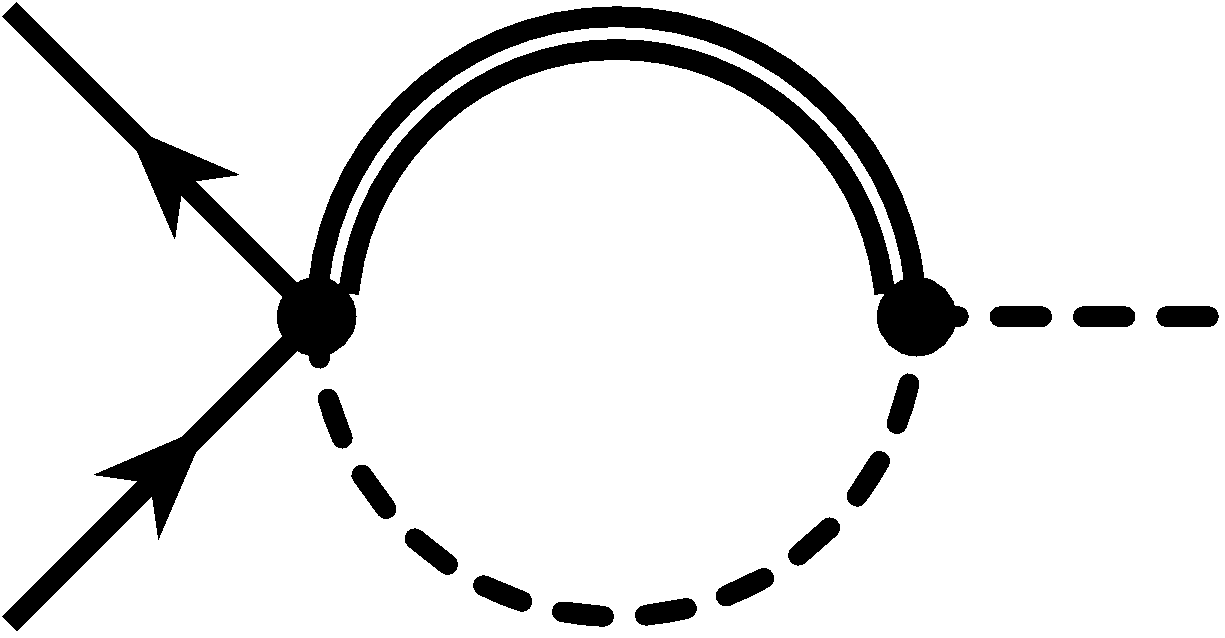}
\hspace{0.02\linewidth}\newline
\includegraphics[width=0.25
\linewidth]{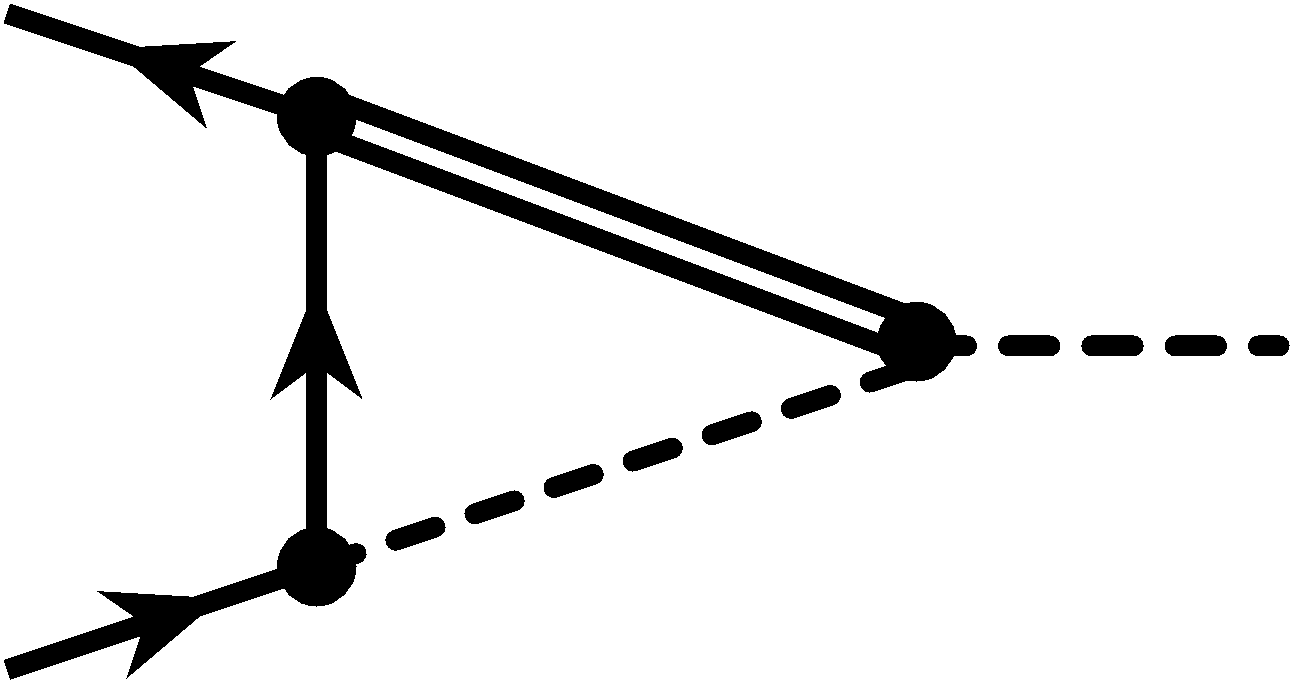}\hspace{0.02\linewidth}
\includegraphics[width=0.25\linewidth]{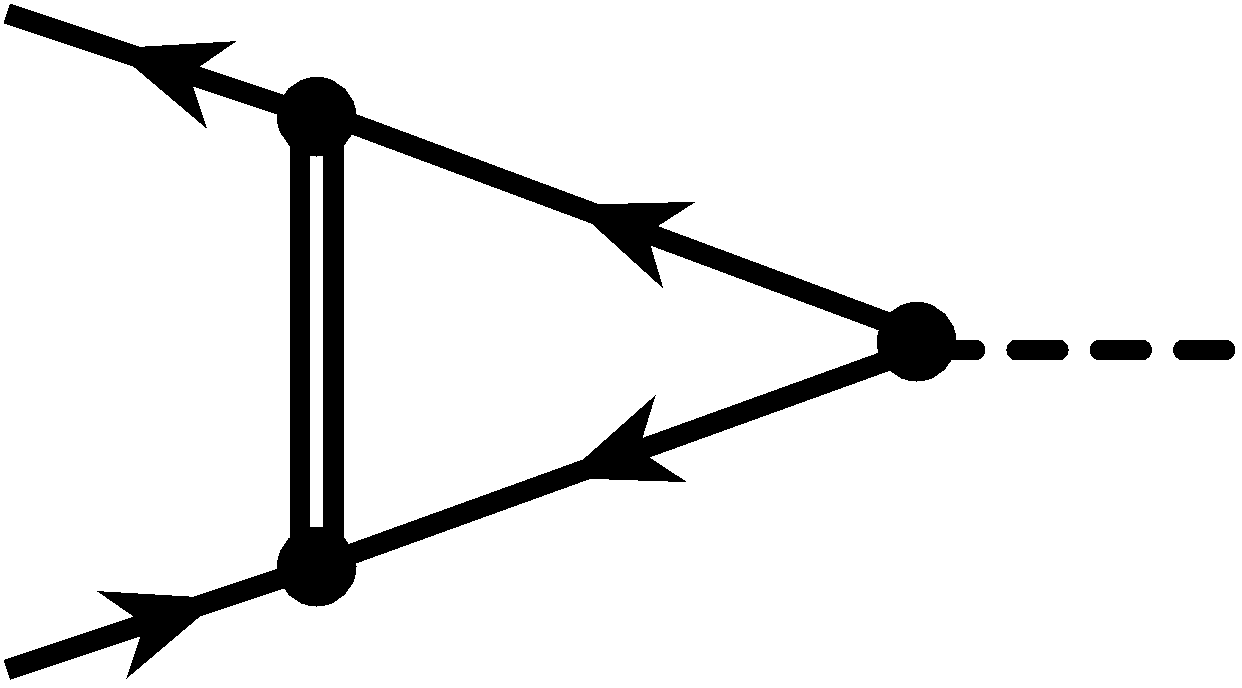}\hspace{0.02\linewidth}
\caption{\label{yukawadiags} Depicting metric propagators by a double
  line, scalars by a dashed line, and fermions by a directed line, we
  show the diagrams contributing to $\beta_y$ that contain at least
  one metric propagator. All diagrams with $n$ different internal
  propagators exist in $n$ versions, with a regulator insertion on
  each of the $n$ interal propagators. All contributions are $\sim g
  y$, and thus contribute directly to the critical exponent at a fixed
  point.}
\end{figure}

We analyze the RG flow of the Yukawa coupling within the truncation
specified by \eqref{eq:truncation} with $\Gamma_{k\, \rm ho}=0$,
cf.~Fig.~\ref{yukawadiags}.
The full result for the gravity-induced part is lengthy and thus
presented in App.~\ref{app:gauge-beta-yukawa}.  We recover the
standard pure-matter contribution that agrees with FRG results in
\cite{Gies:2009hq}, such that
\begin{align} \beta_y =&\, \frac{y}{2}\left(\eta_{\phi,0}+2 \eta_{\psi,0}
  \right)
      \notag\\[2ex]
      & +\frac{y^3}{16 \pi^2}\left(
    \frac{1-\eta_{\psi}/5}{1+\mu_{\phi}}+\frac{1-\eta_{\phi}
      /6}{\left(1+\mu_{\phi}\right)^2}\right) + \beta_{y}^{\rm grav}\,,
\label{eq:betay} 
\end{align} 
where $\beta_{y}^{\rm grav} \sim y \, g$ is proportional to the
dimensionless Newton coupling $g/k^2=G$. It also depends on the
dimensionless graviton mass parameter $ \mu_h k^2=-2\Lambda$, see
App.~\ref{app:gauge-beta-yukawa} for the full expression. In
\eq{eq:betay} we have also used the dimensionless mass parameter of
the scalar field, $\mu_\phi k^2 = m_\phi^2$. Further details on the
graviton propagator can be found in App.~\ref{app:gravity}. Note that
we use that term in a lose way for the propagation of metric
fluctuations. The latter are \emph{not} restricted to small
fluctuations around some background, and the ansatz therefore goes
beyond the perturbative notion of a graviton propagator.

To compare to previous results, we note that if we fix the gauge
parameters to $\alpha=0$ and $\beta=1$, set the graviton mass
parameter and all anomalous dimensions to zero such as to compare to
\cite{Zanusso:2009bs} and \cite{Oda:2015sma}, we come to agreement
with the results obtained in \cite{Oda:2015sma},
\begin{align}
  \beta_y = \frac{y^3(2+\mu_{\phi})}{16\pi^2(1+\mu_{\phi})^2} +g\,y
  \frac{(29-2\mu_{\phi}+5\mu_{\phi}^2)}{20\pi(1+\mu_{\phi})^2}\,.
\end{align}
Note that in their notation $\mu_\phi = 2\lambda_{2}$.  As the term
$y\, \bar{\psi} \psi \phi$ is only symmetric under a combined
$\mathbb{Z}_2$ transformation of the scalar, $\phi \rightarrow -\phi$
and discrete chiral transformation of the fermions, while all other
terms in the truncation are invariant under these transformations
separately, the RG flow of $y$ must be proportional to $y$.  Thus
there is always a Gau\ss ian fixed point $y=0$.  Therefore the
asymptotically safe fixed point that has first been discovered in pure
gravity in \cite{Reuter:1996cp, Reuter:2001ag} and has been shown to
extend to a gravity-matter fixed point in minimally coupled
truncations
\cite{Dona:2013qba,Dona:2014pla,Meibohm:2015twa,Dona:2015tnf}
trivially extends to the case with a Yukawa coupling. In particular,
we can combine the Yukawa coupling with any truncation in the
gravity-matter sector that already features a fixed point, and will
find a trivial generalization of that fixed point.  For definiteness,
let us quote the results obtained in a vertex expansion for gravity
and matter, first analyzed in \cite{Meibohm:2015twa} in the gauge
$\alpha =0,\, \beta=1$, with
\begin{align}
&  g_\ast \approx 0.55,\, \mu_{h,\ast} \approx
  -0.58,\,\eta_{h\,\ast}\approx 0.42,\,\eta_{c,\ast} \approx -1.58,\,
  y_{\ast}=0,\,
  \nonumber\\[2ex]
 & \mu_{\phi\,\ast}=0, \,\eta_{\phi\,\ast}=\eta_{\phi,0\,\ast}
  =0,\,\eta_{\psi\,\ast} \approx 0.07, \,\eta_{\psi,0\,\ast} \approx
  0.72\,,
\label{eq:JJMFP}
\end{align}
with a UV-irrelevant Yukawa coupling. We distinguish the anomalous dimensions from the definition of the vertex, $\eta_{\psi/\phi,\,0}$, from those arising from loop integrals, since our vertices are evaluated at vanishing momentum \cite{Meibohm:2015twa}.
Note that corrections from
Yukawa-diagrams to the matter anomalous dimensions $\eta_{\phi/\psi}$
vanish at a fixed point with $y=0$. Therefore the system in
\cite{Meibohm:2015twa} does not change under the inclusion of a Gau\ss
ian Yukawa-coupling. Hence, by combining the results, we obtain
the critical exponent of the Yukawa coupling,
\begin{align}
  \theta_y = - \frac{\partial \beta_y}{ \partial y} \Big|_{y=0,
    g=g_{\ast}...}\approx- 2.33.\label{eq:thetaJJM} \end{align}
We compare this to the result that can be obtained in a hybrid
background calculation, where the graviton anomalous dimension is
distinguished from that of the background Newton coupling, but the
graviton mass parameter is equated to the background cosmological
constant, \cite{Dona:2013qba}. In that case, we obtain $\theta_y =
-0.06$. While the sign is in agreement with \eqref{eq:thetaJJM},
the absolute value differs significantly. We can trace  that
difference back to the background-approximation for the graviton mass
parameter. In fact, using the fixed-point value for the Newton
coupling and the anomalous dimensions from \cite{Dona:2013qba} and
combining them with the mass parameter from \cite{Meibohm:2015twa},
yields a critical exponent of $\theta_y = -4.88$.

\subsection{Gauge dependence}
Next, we analyze the gauge-dependence of our result by varying the two
gauge parameters $\alpha$ and $\beta$. Note that in a consistent
treatment we would substitute $\alpha\to Z_\alpha \alpha, \, \beta\to
Z_\beta \beta$ in the gauge fixing term \eq{eq:gauge-condition}, see App~\ref{app:gravity}. Here,
we approximate $Z_{\alpha}=1 = Z_{\beta}$.

\begin{figure}[t]
	\begin{center}
          \includegraphics[width=0.43\textwidth]{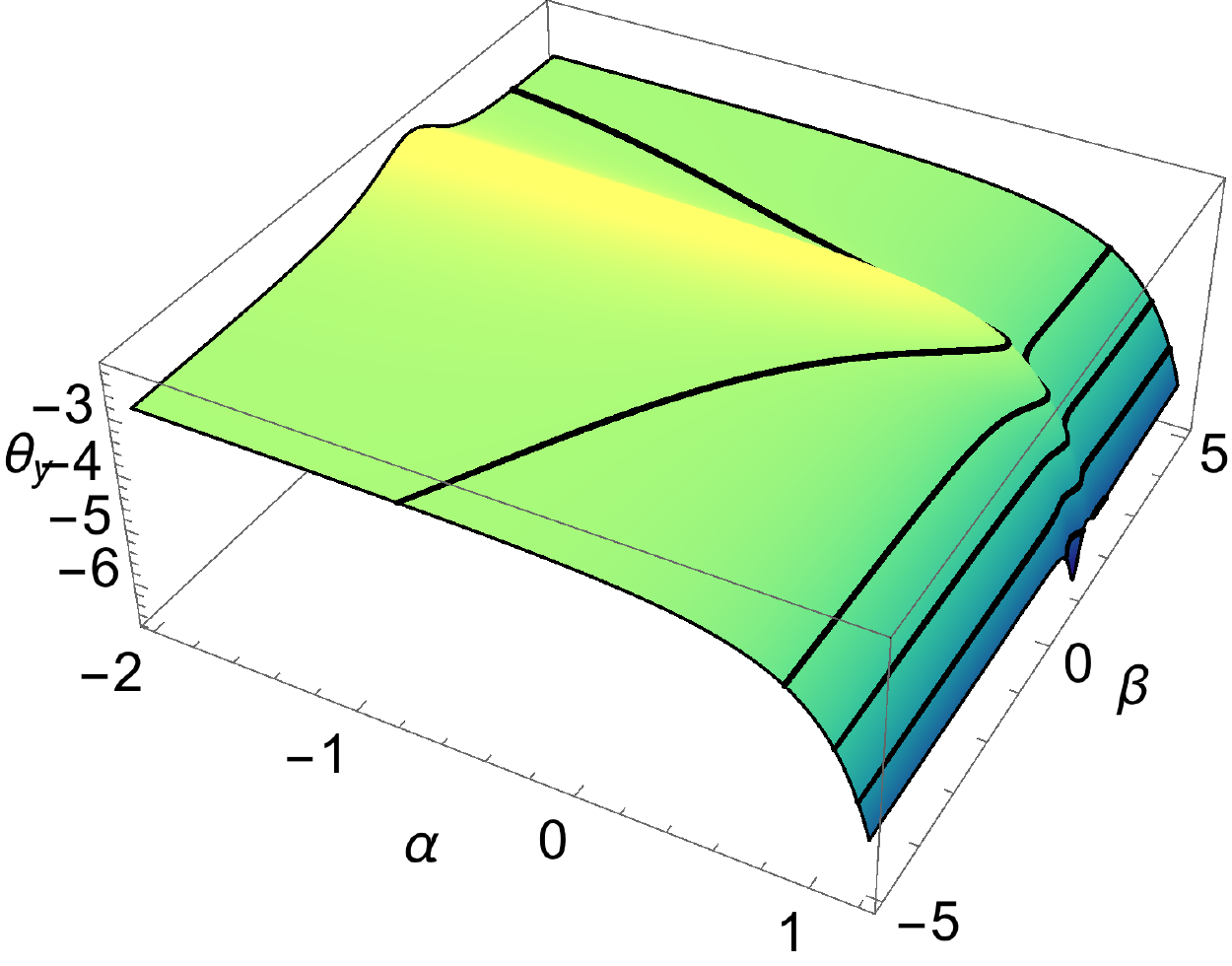}
	\end{center}
	\caption{We plot the critical exponent $\theta_y$ for the
          fixed-point values given in \eqref{eq:JJMFP}
            which were obtained with $\alpha=0$, $\beta=1$. 
          We extrapolate to other values
          of $\alpha$ and $\beta$ without adjusting the fixed-point
          values in the gravitational sector.  }
	\label{fig:gaugeDep-pole-3d-JJM}
\end{figure}

\begin{figure}[t]
\includegraphics[width=\linewidth]{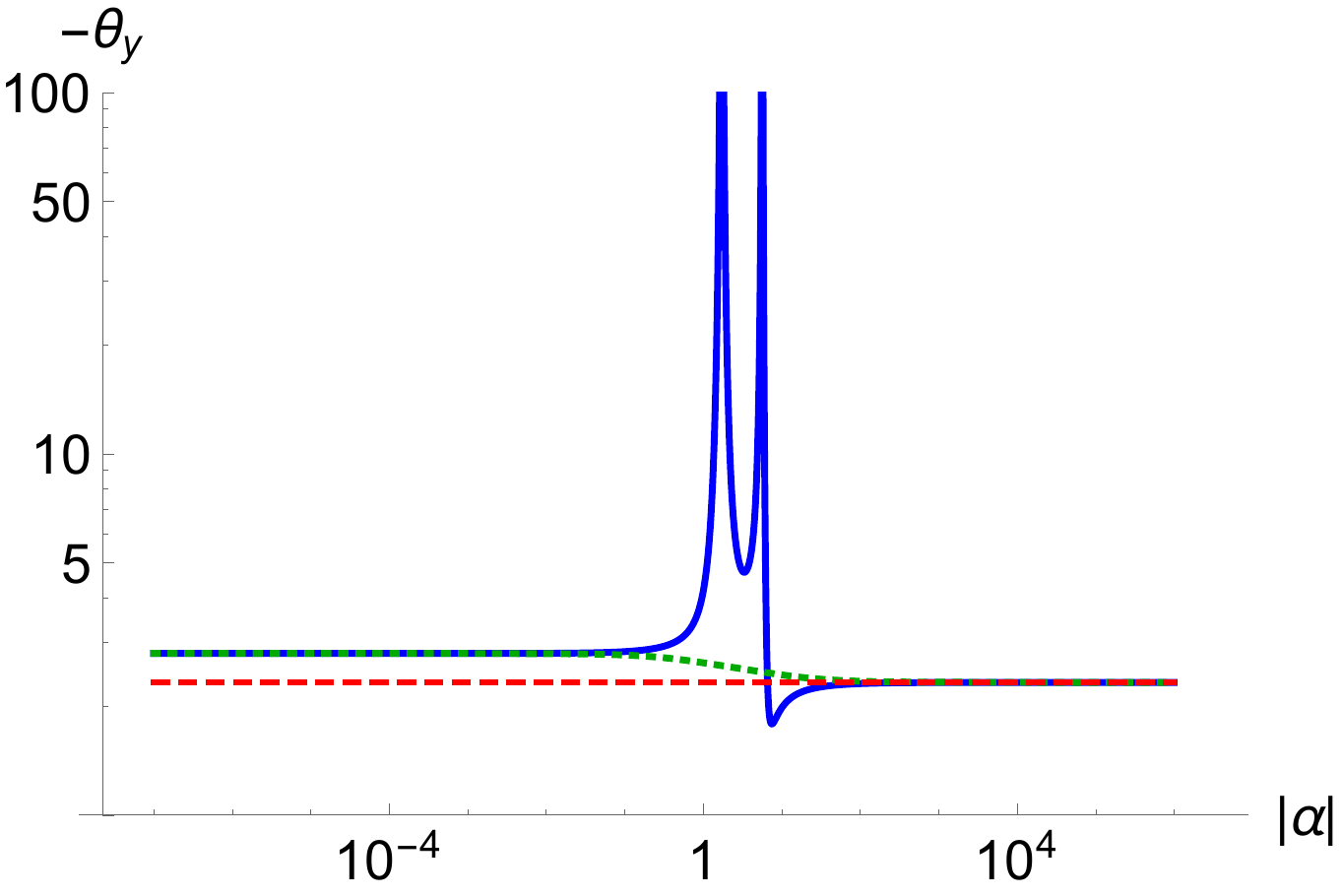}
\caption{\label{fig:theta_y_gauge_JJM} We plot $- \theta_y$ as a
  function of $\alpha$ for the fixed-point values in
  \eqref{eq:JJMFP}. The blue thick curve is for positive $\alpha$,
  while the green dotted one is for negative $\alpha$. At large
  values, both reach the asymptotic limit $\theta_y = -\frac{18}{25} -
  \frac{206838269}{40924800 \pi}$ which is independent of
  $\beta$. The value of $\theta_y$ at $\alpha=0$ depends on $\beta$,
  but is negative for all $\beta$ for the values in
  \eqref{eq:JJMFP}.}
\end{figure}

\begin{figure}[t]
  \includegraphics[width=0.45\linewidth]{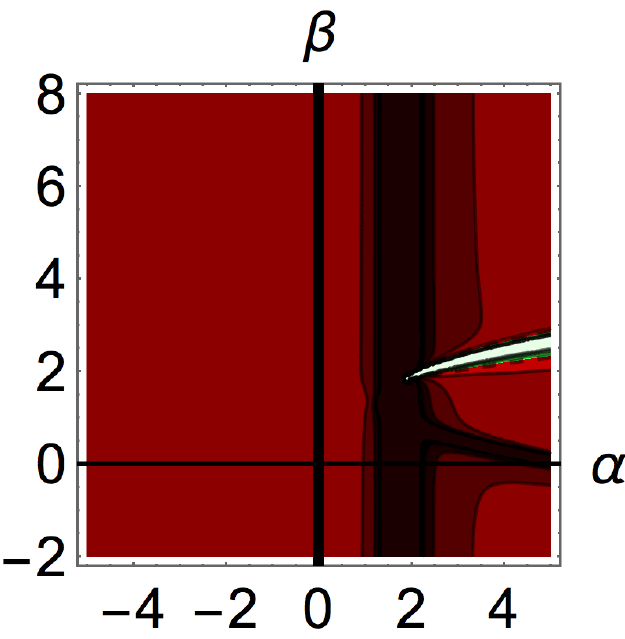}\quad
  \includegraphics[width=0.45\linewidth]{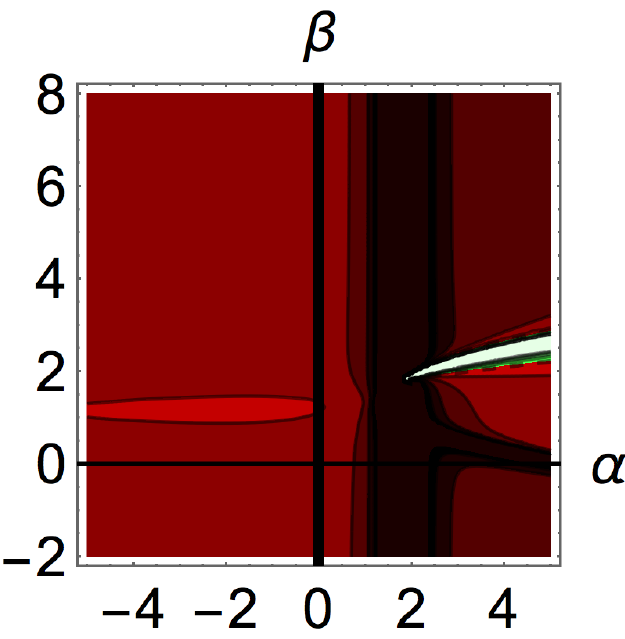}
  \caption{ We plot the critical exponent $\theta_y$ as a function of
    the gauge-parameters $\alpha$ (horizontally) and $\beta$
    (vertically). Green and lighter (red and darker) areas indicate
    positive (negative) values of $\theta_y$ and thus a relevant
    (irrelevant) Yukawa coupling. Contours are drawn at -8, -6, -4,
    -2, 0, 2, 4, 6, 8 and 10. 
    The black axis indicates the line along which 
    the gauge-parameters take fixed
    point values at $\alpha^\star=0$. Additional parameters $(
    g,\mu_h,\mu_\phi,\eta_h,\eta_\phi,\eta_\psi)$ are set to the
    values from \cite{Meibohm:2015twa} (left panel), and from
    \cite{Dona:2013qba} with the fixed-point value for the graviton
    mass parameter taken from \cite{Meibohm:2015twa} (right panel).}
	\label{fig:yukawa-gaugeDep-pole-JJM_mmupgrade}
\end{figure}

\begin{figure}[t]
\includegraphics[width=0.45\linewidth]{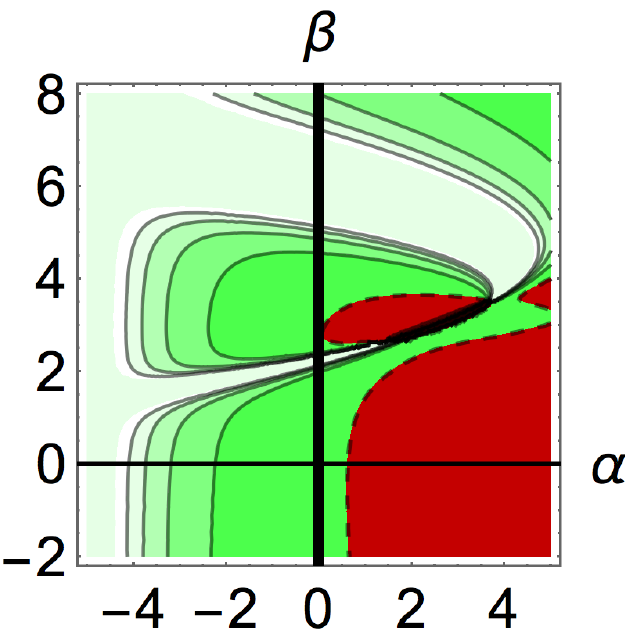}\quad
		\includegraphics[width=0.45\linewidth]{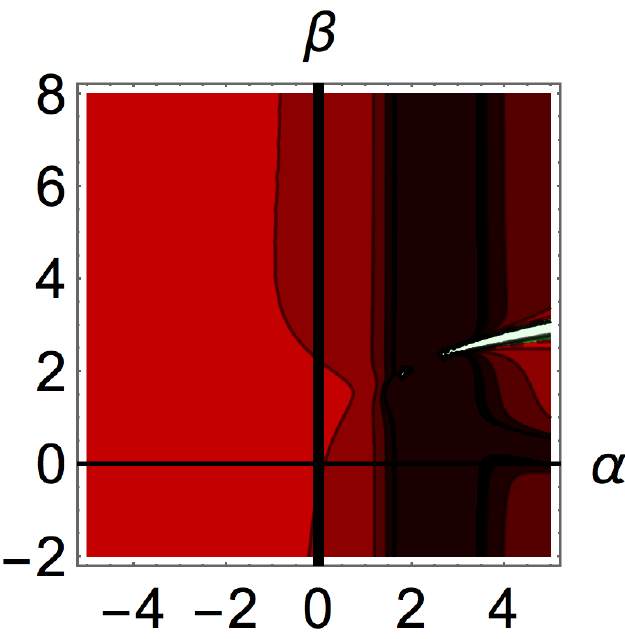}\\
		\includegraphics[width=0.45\linewidth]{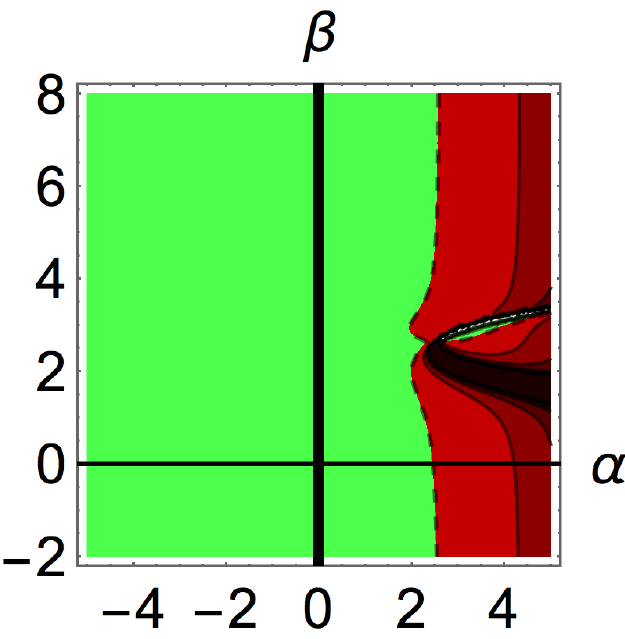}\quad
		\includegraphics[width=0.45\linewidth]{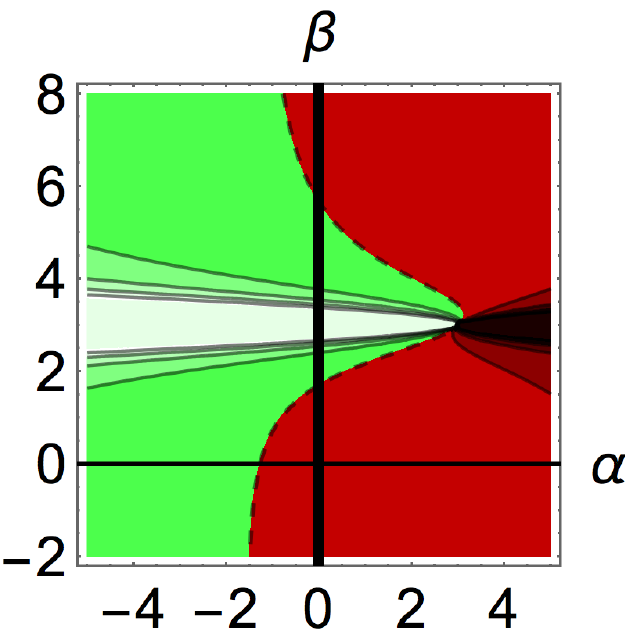}
                \caption{  We plot the critical exponent
                    $\theta_y$ as a function of the gauge-parameters
                  $\alpha$ (horizontally) and $\beta$
                  (vertically). Green and lighter (red and darker)
                  areas indicate positive (negative) values of
                  $\theta_y$ and thus a relevant (irrelevant) Yukawa
                  coupling. Contours are drawn at -8, -6, -4, -2, 0,
                  2, 4, 6, 8 and 10. The dashed black curve highlights
                  the crossing from negative to positive values. The
                  black axis indicates the line along which the
                  gauge-parameters take fixed point values at
                  $\alpha^\star=0$. Additional parameters $(
                    g,\mu_h,\mu_\phi,\eta_h,\eta_\phi,\eta_\psi)$ are
                  set to the values from \cite{Dona:2013qba} (upper
                  left panel), $(0.7,-0.42, 0,-2,0,0)$
                  (cf.~\cite{Becker:2014qya}, upper right panel),
                  $( 0.55,-0.1, 0,-1,-1,0)$ (lower left panel)
                  and $( 0.55,0,0,0,0,0)$ (lower right panel).
                  }
	\label{fig:yukawa-gaugeDep-pole}
\end{figure}

First, we focus on the specific gravitational fixed-point values
quoted in \eqref{eq:JJMFP} in
Figs.~\ref{fig:gaugeDep-pole-3d-JJM}~and~\ref{fig:theta_y_gauge_JJM},
which were obtained with $\alpha=0,\, \beta=1$, but which we
extrapolate to other values of the gauge parameters. We observe poles
at specific values of $\alpha, \beta$, which are induced by an
incomplete gauge-fixing. Extrapolating the fixed-point values for the
gravitational couplings that were obtained for the choice $\alpha=0,
\, \beta=1$ into that region is inconsistent. We expect that in a
complete study, in which the gravitational couplings, all anomalous
dimensions and $\theta_y$ are evaluated in a gauge-dependent way,
these pole-structures disappear. It is reassuring to note that all
values of $\alpha$ and $\beta$ excepting the poles lead to the same
sign for the critical exponent, $\theta_y<0$,
cf.~Fig.~\ref{fig:theta_y_gauge_JJM}.  Within the flow equation,
positive and negative values of $\alpha$ seem admissable, as both
result in an invertible two-point function. On the other hand,
choosing $\alpha<0$ flips the sign of the contribution of the vector
mode to $\beta_y$. Within the path-integral, only $\alpha \geq 0$
naively corresponds to an implementation of the gauge-condition. For
our choice of gravitational couplings, both signs of $\alpha$ lead to
a negative critical exponent for the Yukawa coupling,
cf.~Fig.~\ref{fig:yukawa-gaugeDep-pole-JJM_mmupgrade}. We observe a
similar behavior if we use the fixed-point values from
\cite{Dona:2013qba}, combined with the fixed-point value $\mu_{h\,
  \ast }\approx-0.58$ from \cite{Meibohm:2015twa},
cf.~Fig.~\ref{fig:yukawa-gaugeDep-pole-JJM_mmupgrade}. Resolving the
background-approximation for the graviton mass parameter has a
significant impact on the results, see upper left panel in
Fig.~\ref{fig:yukawa-gaugeDep-pole}: If the fixed-point value for the
background cosmological constant is used for the graviton mass
parameter, the sign of the critical exponent is not stable with
respect to variations of the gauge. Note that in
Fig.~\ref{fig:yukawa-gaugeDep-pole} we use fixed-point values obtained
for the gauge $\beta=1=\alpha$ away from that point. A more consistent
study of the gauge dependence of our result should also include
gauge-dependent gravitational fixed-point values. Nevertheless the
strong gauge dependence could be interpreted as a hint that a
resolution of the background-approximation is of particular importance
for the graviton mass parameter, at least when matter fields are
present.

For vanishing masses and anomalous dimensions the gravitational
contribution to the beta function for the Yukawa coupling reduces to
\begin{align}
\label{eq:betaY-m0eta0}
\beta^\text{grav}_y &=& y\, g\frac{ \left(5 \alpha \left(5 \beta^2-30
      \beta+53\right)+35 \beta^2-258 \beta+339\right)}{20 \pi
  (\beta-3)^2} \,.
\end{align}
For this case, the dependence on the gauge-parameters is depicted in
the lower right panel of Fig.~\ref{fig:yukawa-gaugeDep-pole}. In
\eqref{eq:betaY-m0eta0} a pole occurs in the propagator, when the
gauge-fixing condition \eqref{eq:gauge-condition} does not
correspond to a proper gauge fixing, cf.~Eq. \eqref{eq:h-prop}, as
well as \cite{Gies:2015tca}. While a particular choice of the gauge
parameters can lead to a change of the sign of the critical exponent,
cf.~Fig.~\ref{fig:yukawa-gaugeDep-pole}, it seems to be induced by the
pole in the propagator that arises from a bad choice of gauge,
$\beta=3$.

Varying all gravitational parameters, we observe that some choices of
$\alpha, \, \beta$ can lead to a change in the sign of the critical
exponent, which we tentatively consider an artifact.

\subsection{Stability \& predictive power}
\label{sec:stability}

In the last Section we have utilized results for the Newton coupling
and the graviton mass parameter from
\cite{Meibohm:2015twa,Dona:2013qba} for matter-gravity systems. These
results were obtained in an approximation in which matter has no
self-interactions at the UV fixed point.

Next, we consider effects beyond our truncation which will change the
flow in the gravitational sector and the anomalous dimensions, but
which do not couple directly into $\beta_y$. Some of these will change
the fixed-point values of, e.g., $g$ or $\eta_h$. Thus we test how
strongly any of the parameters $g,
\mu_h,\mu_\phi,\eta_h,\eta_c,\eta_\phi,\eta_\psi$ have to deviate from
the results in \eqref{eq:JJMFP} for the Yukawa coupling to become
relevant.

\begin{figure}[!t]
\includegraphics[width=\linewidth]{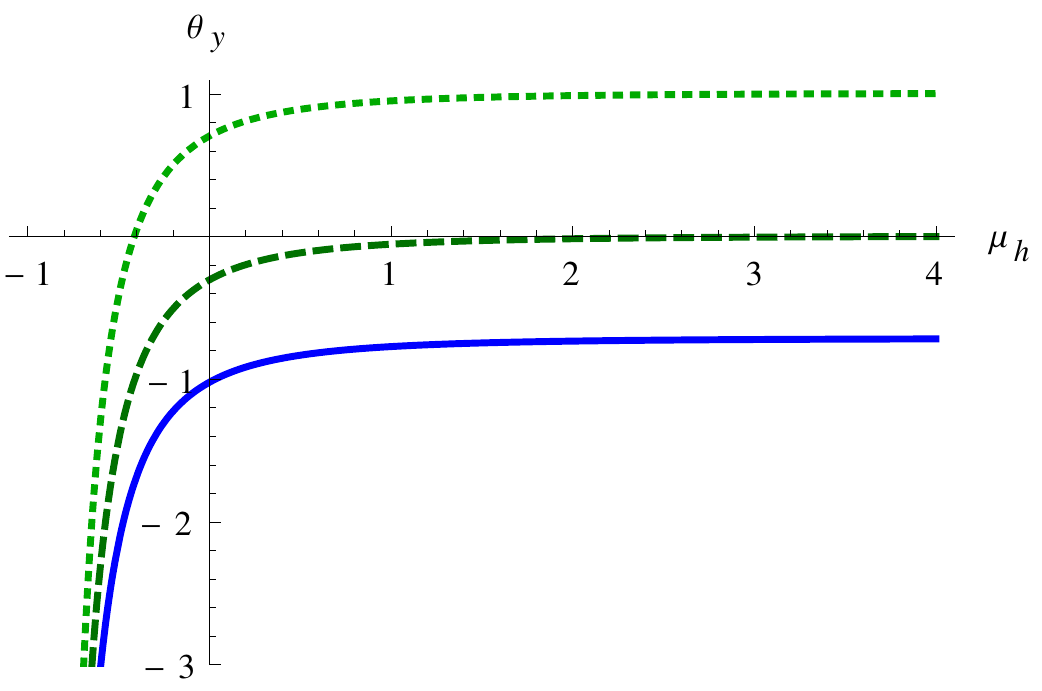}
\caption{\label{yukawa_theta_gravmass} We plot the gravitational
  contribution to the critical exponent of the Yukawa coupling divided
  by the Newton coupling, for gauge parameters $\beta=\alpha=0$, and
  $g^\ast,\eta_h^\ast$ from \cite{Meibohm:2015twa}, as a function of
  the graviton mass parameter $\mu_h$ for $\eta_\psi = 0.07$ \&
  $\eta_{\psi,0} =0.72$ (cf. \eqref{eq:JJMFP}; thick blue line),
  $\eta_\psi=\eta_{\psi,0}=0$ (dark green dashed line) and
  $\eta_\psi=\eta_{\psi,0}=-1$ (light green dotted line). }
\end{figure}

We then focus on the preferred choice of gauge $\alpha =0$, which
corresponds to an RG fixed point \citep{Litim:1998nf}. The latter
holds, as the gauge-fixing condition is strictly imposed, which cannot
be changed by the -finite- flow. Further, we choose $\beta = \alpha $,
while noting that the choice $\beta=1$, $\alpha=0$ does not lead to
any qualitative differences in our results, cf.~Sec.~\ref{sec:gauge}.
We observe that a large positive mass-parameter of the
graviton -- corresponding to a negative cosmological constant in a
single-metric truncation or a negative level-2-cosmological constant
in a ``bi-metric" truncation -- can change the sign of the critical
exponent, cf.~Fig.~\ref{yukawa_theta_gravmass}, as can negative
anomalous dimensions for matter. For instance, setting $\mu_{\phi}=0$
leads to
\bea \theta_y &:=& - \frac{\partial \beta_{y}}{\partial_y}\Big|_{y =0}
= - \frac{1}{2} (\eta_{\phi,0}+ 2 \eta_{\psi,0} ) +g \frac{231
  - 41 \eta_{\psi}}{280 \pi (3+2 \mu_h)} \nonumber\\
&{}&- \frac{g}{6720 \pi} \left(\frac{2800 (6- \eta_h)}{(1+ \mu_h)^2} -
  3\frac{6888 + 869 \eta_h}{(3+ 2 \mu_h)^2}
\right).\label{thetaYanomdim} \eea
At large enough $\mu_h$ and/or large negative $\eta_{\phi/\psi}$, the
last line in \eqref{thetaYanomdim} is suppressed, and the positive
contribution dominates.  In that case, two further fixed-point
solutions are pulled from the complex plane onto the real axis.
The non-Gau\ss ian fixed-point values are given by
\begin{widetext}
  \begin{align}
 y_{\ast}=\;\pm &&\sqrt{\frac{\pi}{14(-6 \eta _{\psi }-5 \eta
      _{\phi }+60)}} \sqrt{g\left(\frac{24 \left(231-41 \eta _{\psi
          }\right)}{3+2\mu _h}+\frac{3 \left(6888-869 \eta
          _h\right)}{\left(3+2\mu_h\right){}^2}-\frac{2800
        \left(6-\eta _h\right)}{\left(1+\mu_h\right){}^2}\right)-3360
    \pi \left(2 \eta _{\psi,0}+\eta _{\phi,0}\right)}.  
    \label{eq:yStar}
    \end{align}
    \end{widetext}
    We emphasize that large positive values of $\mu_h$ are far from
    being accessible in state-of-the-art truncations in
    gravity. Nevertheless, they might be realized in extended
    truncations.

    While matter interactions do not contribute directly to the
    pure-gravity parameters, they couple back into the matter
    anomalous dimensions $\eta_{\phi/\psi}$. Even for a vanishing
    Yukawa coupling, tadpole diagrams from matter 4-point interactions
    will contribute, cf.~Sec.~\ref{sec:quartic}. The corresponding
    changes in the anomalous dimensions could present another possible
    mechanism for pushing the Yukawa coupling into relevance, as
    negative (positive) $\eta_{\phi/\psi}$ tend to make the Yukawa
    coupling more relevant (irrelevant),
    cf.~Fig~\ref{yukawa_theta_gravmass}. Perturbing the fixed-point
    values from \eqref{eq:JJMFP}, the case of $\eta_\psi=
    \eta_\phi=-1.411$ features a real fixed point
    (cf.~\eqref{eq:yStar}) at $y_{\ast}\approx \sqrt{2}\cdot 0.4$ with
    critical exponents $\theta_y=-0.01$.  This would enforce that
    particular value of the Yukawa coupling in the vicinity of the
    Planck scale, corresponding exactly to the value of the top-Yukawa
    coupling at that scale.  Note that this scenario is contingent
    upon large negative values for the anomalous dimensions, which
    would require substantial changes from the results in
    \eqref{eq:JJMFP}. In general a scenario with additional non-Gau\ss
    ian fixed points in the Yukawa coupling can be realized whenever
    there are negative contributions to the Yukawa
    $\beta$-function. In the present truncation, supplemented by
    gravitational fixed-point values from \eqref{eq:JJMFP}, such
    contributions are absent.

We conclude that in the present simple truncation the Yukawa coupling
exhibits a Gau\ss ian fixed point with  irrelevant
UV-behavior. We also observe that our calculation is compatible with a variety of
approximations in the metric sector. For instance, within a
single-metric approximation, $g$ would correspond to the
background-value for the Newton coupling, and $\mu_h = -2 \Lambda$. As
an example, we use fixed-point values from \cite{Gies:2015tca} with
$\eta_h=-2$, which yield $\theta_y=-1.44$ for $g_{\ast}=0.89,\,
\mu_h=-0.33$, $\alpha=0, \, \beta=1$, and $\theta_y=-1.97$ for
$g_{\ast}=0.88,\, \mu_h =-0.36$ for $\beta=0=\alpha$ and finally
$\theta_y = -2.34$ for $g_{\ast}=0.72,\, \mu_h=-0.32$ for
$\beta=1=\alpha$.  In a ``bi-metric" calculation, the gravitational
couplings entering the beta function for the Yukawa coupling would be
those of the dynamical metric. Choosing $g_{\ast}\approx0.7,\,
\mu_h\approx -0.42,\, \eta_h=-2$ and $\alpha=1, \beta=1$, as in
\cite{Becker:2014qya} in fact leads to a critical exponent of
$\theta_y=-3.16$.  Thus, our conclusions on the critical exponent of
the Yukawa coupling do not rely on specific assumption for the
approximation in the gravitational sector, but hold
 using fixed-point values from various studies in the
  literature.

  The irrelevance of this coupling signals that its low-energy value
  can be predicted. This could open the door to an observational test
  of asymptotically safe quantum gravity, as the low-energy values of
  the Yukawa couplings in the Standard Model are known
  experimentally. Neglecting the curvature of the critical surface,
  the fixed point cannot be reached in the UV, unless the low-energy
  value of the coupling already equals the fixed-point value. We would
  thus conclude that at the Planck scale, where the joint
  matter-gravity RG flow sets in, the Yukawa coupling would already
  have to vanish in order for the RG trajectory to reach the fixed
  point. While this is close to the right value for the Yukawa
  couplings of the Higgs to the bottom and all lighter quarks and
  leptons, we know that the top-Yukawa coupling in the Standard Model
  is of the order $y \approx 0.4/\sqrt{2}$ at the Planck
  scale. This might suggest that the fixed point that we have analyzed
  here is not one which is compatible with the properties of the
  Standard Model at low energies. Note however, that statements about
  quantum gravity effects on the running couplings of the Standard
  Model are also generically scheme-dependent, see, e.g.,
  \cite{Robinson:2005fj,Pietrykowski:2006xy,Ebert:2007gf,Daum:2009dn,%
    Toms:2010vy,Toms:2011zza,Folkerts:2011jz}. As couplings do not
  directly correspond to observable quantities, non-universality is
  not a problem. Still, it implies that only results obtained in the
  same scheme can be compared. Finally, here we only consider a toy
  model of the Yukawa sector, and have neglected additional matter
  fields of the Standard Model, which have to be included, e.g., along
  the lines of \cite{Gies:2014xha}. In summary, such a comparison
  requires further work in order to be dependable.

Moreover extensions of the truncation are clearly indicated: A
reliable evaluation of the anomalous dimensions for the matter fields
is critical to settle these questions. As has already been pointed out
for scalar fields and fermions
  \cite{Eichhorn:2011pc,Eichhorn:2012va}, quantum gravity
fluctuations induce further momentum-dependent interactions at the
fixed point. These couple directly into the flow of the anomalous
dimension, and are thus of critical importance for our study.
This will entice us to consider in more detail, whether the above
truncation already captures all important effects of quantum gravity
on the Yukawa sector. As we will show in the next sections, it does
not.

%
%
\section{Gravity-induced matter interactions}
\label{sec:quartic}

\begin{figure}[t]
	\begin{center}
		\hfill
		\includegraphics[width=0.1\textwidth]{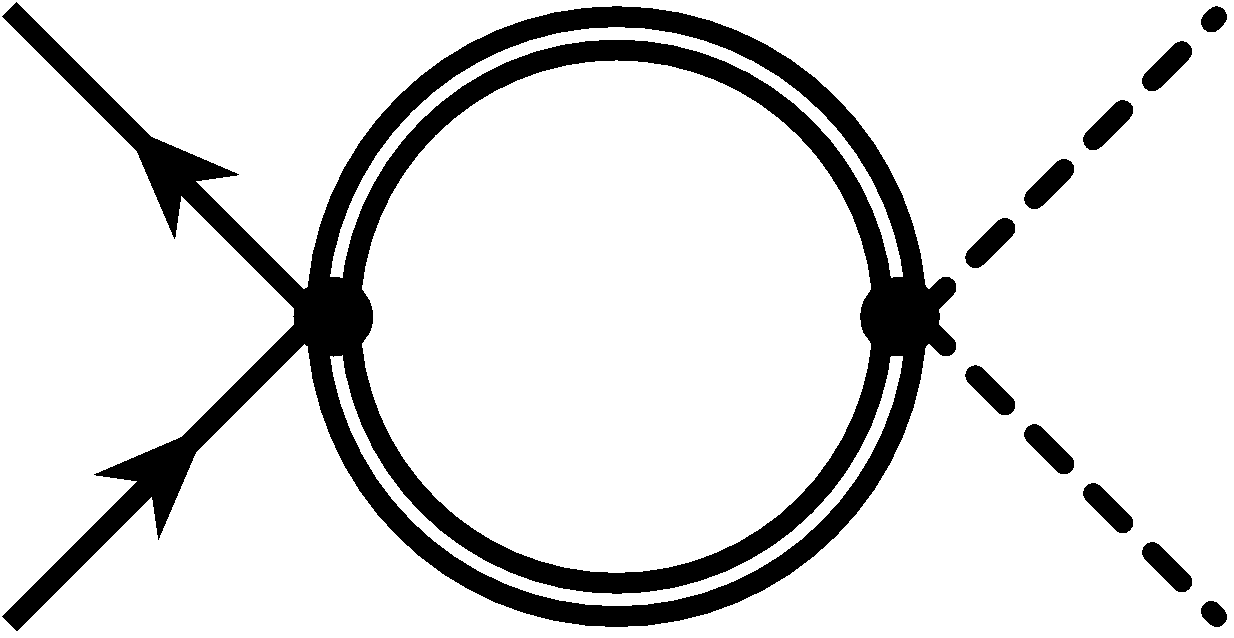}
		\hfill
		\includegraphics[width=0.1\textwidth]{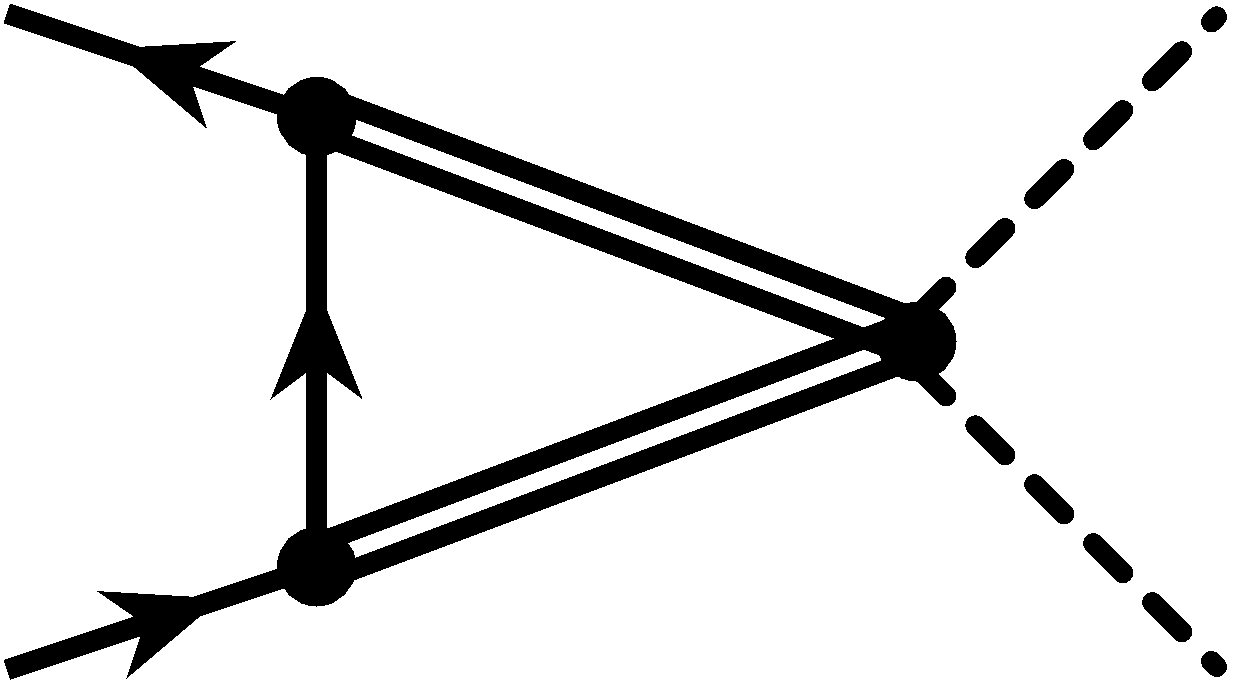}
		\hfill
		\includegraphics[width=0.09\textwidth]{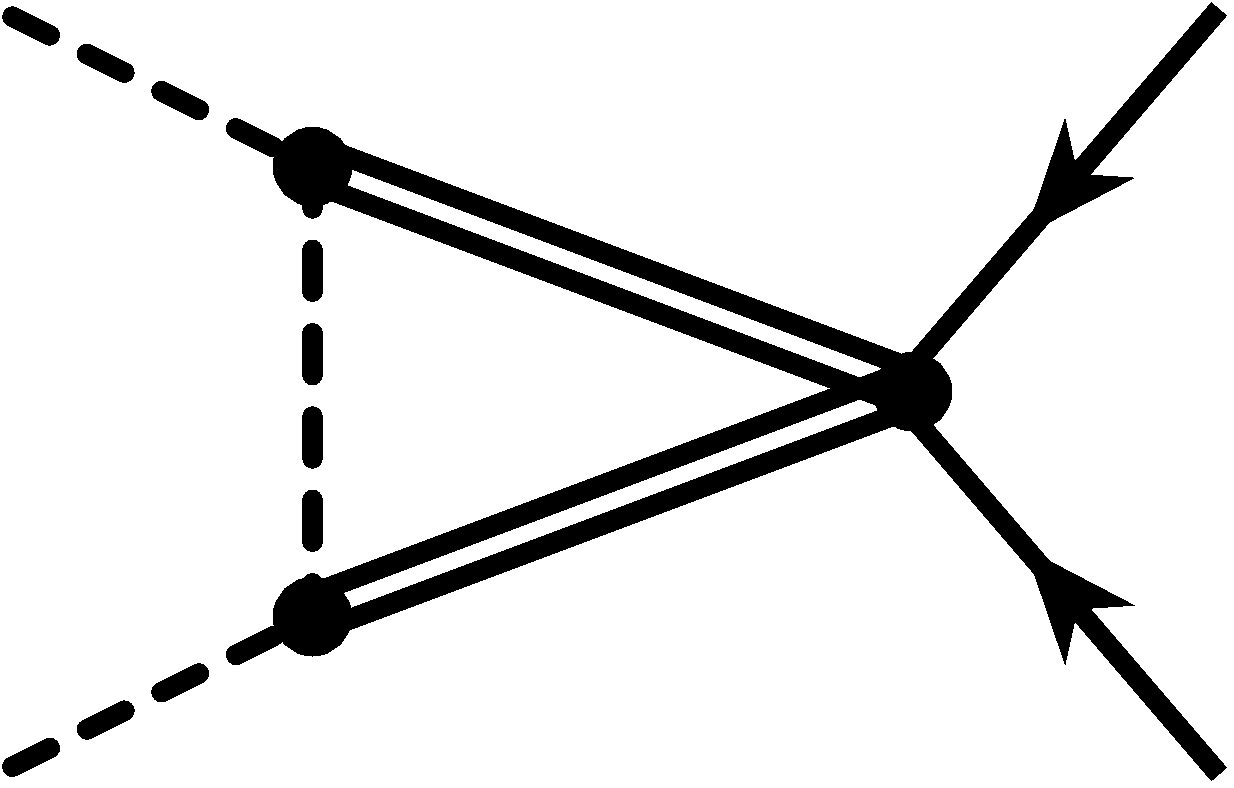}
		\hfill	
		\includegraphics[width=0.1\textwidth]{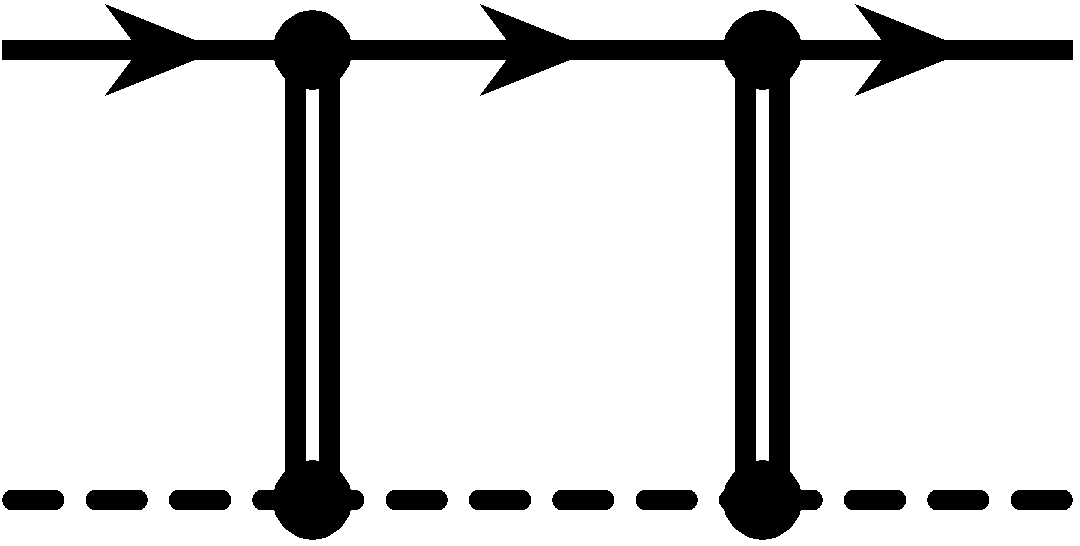}
	\end{center}
	\caption{ Mixed quartic diagrams generated by covariant
          kinetic terms in the flow equation. These diagrams are
          nonzero, even if all matter self-interactions are set to
          zero initially. They prevent the corresponding
          2-scalar-2-fermion couplings from being able to reach
          asymptotic freedom.  }
	\label{fig:quartic-flow-diags}
\end{figure}

As has been pointed out for scalar and fermionic systems separately
\cite{Eichhorn:2011pc,Eichhorn:2012va,Eichhorn:2013ug,Meibohm:2016mkp},
asymptotically safe quantum gravity induces non-vanishing matter
self-interactions at the fixed point,
cf.~Fig.~\ref{fig:quartic-flow-diags}. Thus the only possible fixed
points that exist for the joint matter-gravity system are necessarily
non-Gau\ss ian in some of the matter couplings. Clearly, the Yukawa
coupling is not one of those couplings, as all couplings with an odd
number of matter fields can vanish. This scenario is as close as
possible to a fixed point that is Gau\ss ian in the matter
sector. Then, the first, genuinely gravity-induced couplings are the
matter four-point vertices. Here we investigate the
two-fermion--two-scalar couplings. The four-fermi, and four-scalar
couplings have been discussed separately in gravity-fermion systems,
\cite{Eichhorn:2011pc,Meibohm:2016mkp}, and gravity-scalar systems,
\cite{Eichhorn:2012va}. Both of them couple nontrivially into the flow
of two-fermion--two-scalar couplings. Here, we make a first step in
the exploration of this system by focussing on the mixed interactions,
only.

In that system, there are two main questions:
\begin{itemize}
\item Is there a fixed point in the matter-gravity system?
\item Are canonically irrelevant couplings shifted into relevance at
  the joint fixed point?
\end{itemize}
Note that the first is a vital question since it is a necessary
condition for the realization of asymptotic safety in a quantum theory
of gravity combined with matter fields. 

\subsection{Synopsis}
In this section, we show that certain momentum-dependent
scalar-fermion interactions \emph{cannot} have a Gau\ss ian fixed
point in the presence of asymptotically safe gravity. In particular we
show that the Gau\ss ian matter fixed point for vanishing
gravitational coupling $g$ is shifted to an interacting one at finite
$g$, called the shifted Gau\ss ian fixed point (sGFP). Thus,
asymptotically safe quantum gravity cannot be coupled to a fully
asymptotically free matter system. At least a subset of the matter
couplings must become asymptotically safe.  This entails that the
scaling dimensions of these couplings depart from canonical
scaling. In fact, we observe that some of the canonically irrelevant
scalar-fermion couplings will be pushed towards relevance.

Moreover, we find that for large enough gravitational coupling, the
sGFP moves off into the complex plane, i.e., quantum-gravity effects
can lead to fixed-point annihilations in the matter sector. Then, the
only remaining viable fixed points are fully interacting and exhibit a
larger number of relevant directions. These fixed-point
  annihilations require the effective strength of gravitational
  interactions to exceed a critical value which is well beyond that
  reached in results in the literature, see, e.g.,
  \cite{Becker:2014qya,Meibohm:2016mkp,Dona:2013qba}.

Our main results are summarized in Fig.~\ref{fig:xFPexistencePlane},
where light grey dots mark a region where the shifted Gau\ss ian fixed
point does not exist. The fixed-point values obtained in
\cite{Meibohm:2015twa} lie within the green region, where the
shifted Gau\ss ian fixed point exists. 
\subsection{Induced two-fermion--two-scalar interactions}
From here on, we work in the gauge $\alpha =0$ and $\beta=1$.
Including the covariant matter kinetic terms in the action
automatically generates vertices with two matter fields and an
arbitrary number of gravitons, i.e.,
\begin{eqnarray}
	S_{\psi,\text{kin}} &=& 
	i\int d^4x\sqrt{g}\; \left(\bar{\psi}\gamma_\mu\nabla^\mu\psi\right)
	\label{eq:action_kineticPsi}\\
	\quad &\Longrightarrow & \quad
	\raisebox{-.37\height}{
		\includegraphics[width=0.1\linewidth]{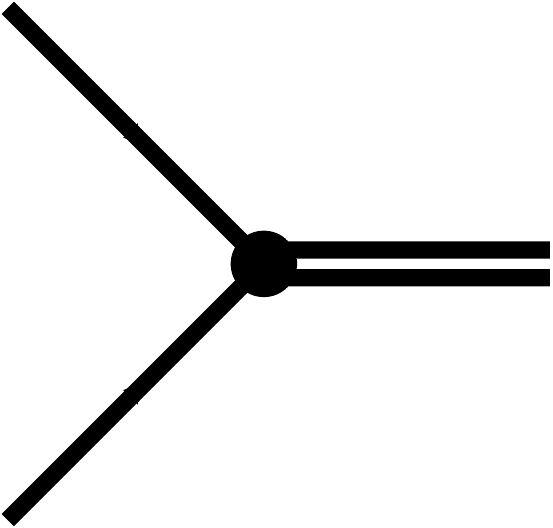}\hspace{0.05\linewidth}
		\includegraphics[width=0.1\linewidth]{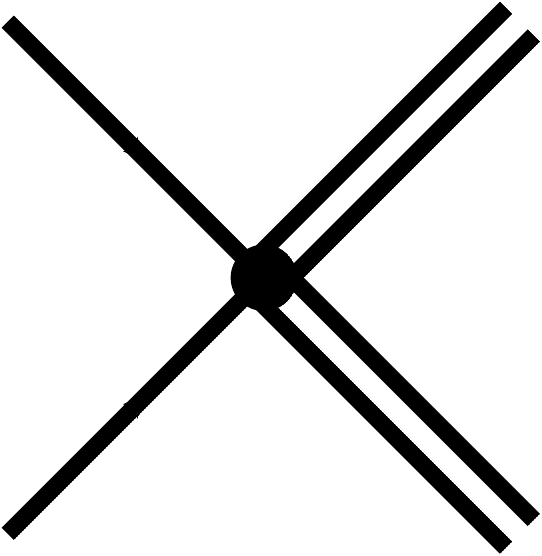}\hspace{0.05\linewidth}
	}
	\cdots
	\quad
	\raisebox{-.37\height}{
		\includegraphics[width=0.11\linewidth]{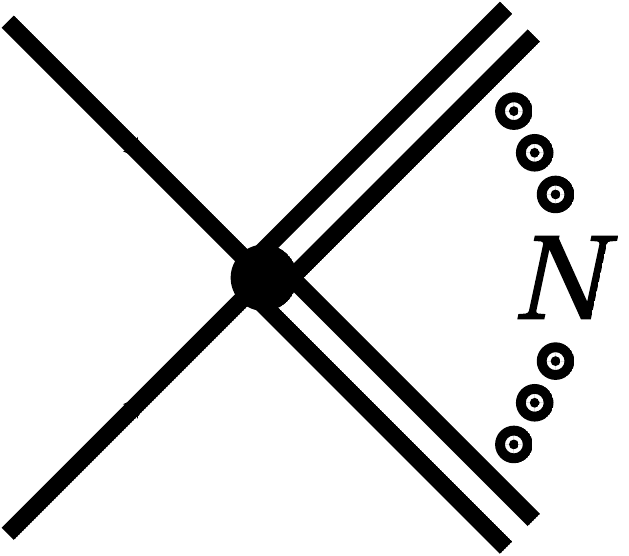}\hspace{0.05\linewidth}
	}
	\notag	
	\\[2ex]
	S_{\phi,\text{kin}} &=& 
	\frac{1}{2}\int d^4x\sqrt{g}\; \left(g^{\mu\nu}\partial_\mu \phi \partial_\nu \phi\right)
	\label{eq:action_kineticPhi}\\
	\quad &\Longrightarrow & \quad
	\raisebox{-.37\height}{
		\includegraphics[width=0.1\linewidth]{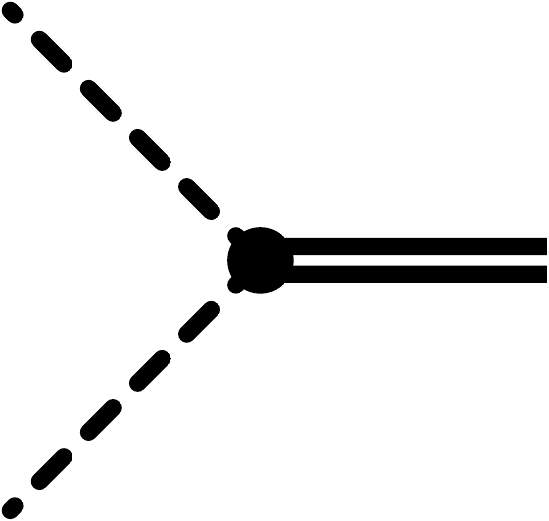}\hspace{0.05\linewidth}
		\includegraphics[width=0.1\linewidth]{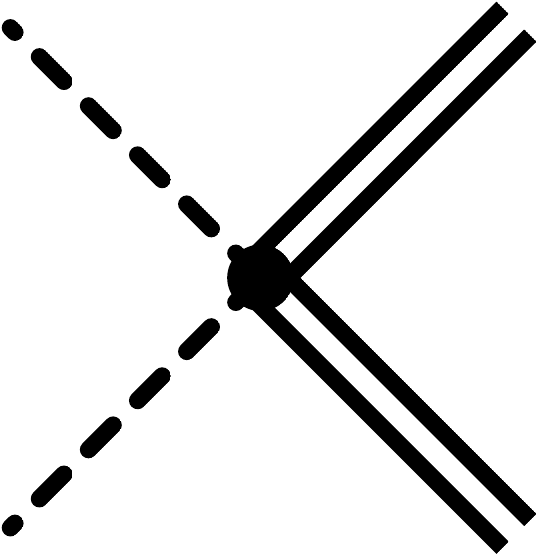}\hspace{0.05\linewidth}
	}
	\cdots
	\quad
	\raisebox{-.37\height}{
		\includegraphics[width=0.11\linewidth]{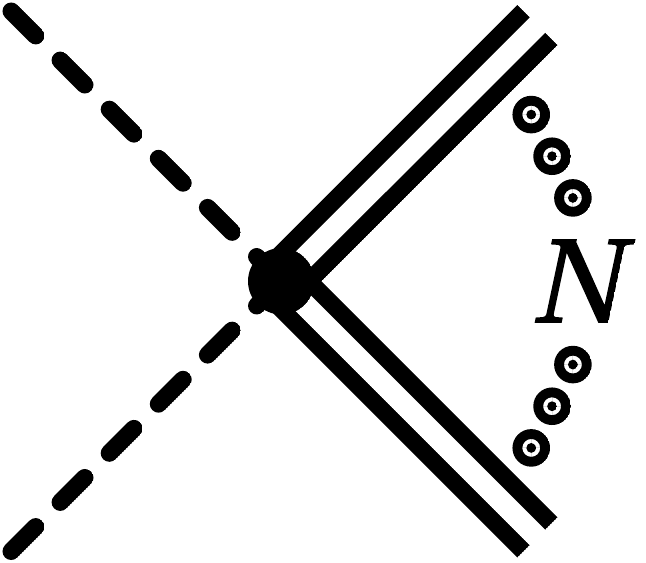}\hspace{0.05\linewidth}
	}
	\notag .	
\end{eqnarray}
These vertices give rise to diagrams that induce higher order
$2s$-point, $2f$-point and $2s$-$2f$-point functions, where $2s\,, 2f$
is the number of external scalar and fermion fields
respectively. Couplings with an odd number of external fermions or
scalars are not purely gravity-induced from the kinetic terms.  We
follow canonical power counting, and focus on the leading-order terms
in the corresponding expansion in the remainder of this section, which
has $s= f= 2$. The fixed-point properties of the corresponding
couplings are encoded in the four types of diagrams shown in
Fig.~\ref{fig:quartic-flow-diags}.  These diagrams are
\emph{non-vanishing}, \emph{even if all matter self-interactions are
  set to zero}. Thus, if we switch off all matter self-interactions,
but keep a finite gravitational coupling, the beta-functions of these
quartic scalar-fermion couplings will be nonzero.  Accordingly, there
are scalar-fermion couplings which \emph{cannot} vanish at a
matter-gravity fixed point. This leads us to the canonically most
relevant induced  structures. These are quartic in the fields and of order
$p^3$.  A complete basis of these operators manifestly obeying
reflection positivity of the Osterwalder-Schrader theorems
\cite{Osterwalder:1973dx, Wetterich:2010ni} is built from the
reflection positive combinations 
\begin{align}\nonumber 
	&\bar\psi\gamma^\mu\partial^\nu\psi + (\partial^\nu\bar\psi)\gamma^\mu\psi \;,
	\\[2ex]
	&i \bar\psi\gamma^\mu\partial^\nu\psi - i (\partial^\nu\bar\psi)\gamma^\mu\psi\;,
	\label{eq:refPosAndEuclidImReal}
\end{align}
traced either into $\delta_{\mu\nu}\partial^\rho\phi\partial_\rho\phi$
or into $\partial_\mu\phi\partial_\nu\phi$. The first
  reflection positive combination in \eqref{eq:refPosAndEuclidImReal}
  corresponds to imaginary flows in Euclidean space. However, these
  operators are neither created by gravity nor by the other reflection
  positive operators. Therefore these imaginary couplings are
  completely decoupled from the system and remain zero. An intuitive
  reasoning for this can be given since all other operators in the
  initial effective action are reflection positive and real in
  Euclidean space and the flow preserves these properties. A detailed
  discussion of this issue can be found in App.~\ref{app:proj}.  The
  reflection positive and Euclidean-space real combinations amount to 
the gravity-induced higher order couplings already put down in
\eq{eq:YukawaX}. For the sake of a fixed-point analysis it is
convenient to introduce dimensionless couplings, to wit
\begin{align}\label{eq:X}
 \mathcal{X}_i =
  \bar{\mathcal{X}}_i k^4.  
\end{align}
Inserting the definition \eq{eq:X} leads us to a complete basis of
induced structures 
\begin{align}\nonumber 
&\0{ \Gamma_{k\, \rm induced}}{Z_{\phi}\, Z_{\psi} { k^4}} = \\[2ex] 
= &\, i\mathcal{X}_{1-} \int d^4x\sqrt{g}\; \left[
  \left(\bar\psi\gamma^\mu\nabla_\nu\psi -
    (\nabla_\nu\bar\psi)\gamma^\mu\psi\right)
  \left(\partial_\mu\phi\partial^\nu\phi\right) \right]
\notag\\[2ex]
+ &\, i\mathcal{X}_{2-}\int d^4x\sqrt{g}\; \left[
  \left(\bar\psi\gamma^\mu\nabla_\mu\psi -
    (\nabla_\mu\bar\psi)\gamma^\mu\psi\right)
  \left(\partial_\nu\phi\partial^\nu\phi\right) \right] \,,
\label{eq:YukawaXre}\end{align}
where the flow of the left hand side can be easily projected on the
couplings $\mathcal{X}_i$, see App~\ref{app:proj}.

\subsection{Fixed-point shifts and annihilations in a simplified
  system}\label{simpletruncation}
\subsubsection{Beta functions}

First we analyze the two distinct interactions in \eqref{eq:YukawaXre} 
grouped together in a single coupling 
\begin{equation}
\mathcal{X} = \frac{1}{2}(\mathcal{X}_{1-} + \mathcal{X}_{2-})
\end{equation}
Therefore we combine the flows obtained with the projection rules
specified in \eqref{eq:projX3m} \& \eqref{eq:projX4m} according to the
above relation and employ $\mathcal{X}_{1-} = \mathcal{X}_{2-} =
\mathcal{X}$. This yields a $\beta$-function for the combined
coupling. The intriguing properties that we will observe in this
simplified system persist within a more extended analysis, see
Sec.~\ref{sec:fully_disentangled_truncation}. For the sake of
simplicity we have set the metric mass parameter and all anomalous
dimensions to zero.  The $\beta$-function for the joint
  coupling $\mathcal{X}$ then reads 
\begin{align} 
  \beta_{\mathcal{X}} =\,4 \mathcal{X}+ \frac{7}{2} \,g^2 -
  \frac{99}{160 \pi}g\, \mathcal{X}
  \,+ \frac{143}{896 \pi^2} \mathcal{X}^2\,,
 \label{eq:betax}\end{align}
The key effect of gravity is encoded in the term $7/2\, g^2$ in
Eqs.~\eqref{eq:betax}, which implies that no Gau\ss ian fixed point in
$\mathcal{X}$ can exist at finite $g$.  Thus, a major result of ours
is that as soon as metric fluctuations are switched on, the Gau\ss ian
fixed point is shifted to become interacting
(cf.~Fig.~\ref{fig:betaX}).  Accordingly, its scaling exponent
deviates from the canonical value $-4$, and becomes less irrelevant.
At finite $g$, this fixed point is distinct from other possible
interacting fixed points as its scaling exponent still reflects the
canonical irrelevance of the coupling.

\subsubsection{Fixed-point annihilations and excluded regions}
\begin{figure}[t]
\includegraphics[width=\linewidth]{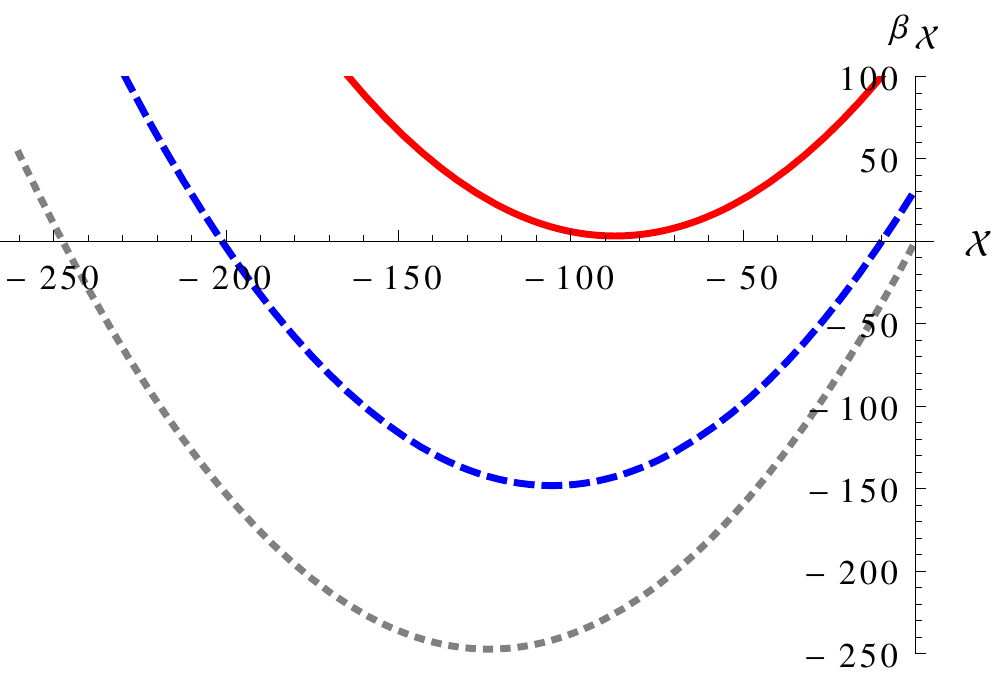}
\caption{\label{fig:betaX} We plot the beta-function for $\mathcal{X}
  = (\mathcal{X}_{1-} + \mathcal{X}_{2-})/2$ for vanishing anomalous
  dimensions and metric mass parameter and for $g=0$ (gray dotted
  line), $g=3$ (blue dashed line) and $g\equiv g_\text{crit.} \approx
  6$ (red line). The Gau\ss ian fixed point at $\mathcal{X}=0$ only
  exists for $g=0$, and then becomes shifted to an interacting fixed
  point at finite g, which annihilates with the second interacting
  fixed point at $g_\text{crit.} \approx 6$.}
\end{figure}

At finite gravitational couplings $g, \mu_h$, the system shows an
intriguing fixed-point behavior, because the non-Gau\ss ian
gravitational contributions, i.e., the second and third term in
Eq.~\eqref{eq:betax}, are potentially destabilizing. This is best
understood starting from the gravity-free case: At $g=0$, the system
features a single Gau\ss ian fixed point, cf.~\eq{eq:betax}.  As a
function of $g$, this fixed point moves away from the origin,
cf.~Fig.~\ref{fig:betaX}.  Simultaneously, the non-Gau\ss ian
  fixed point moves towards the origin, such that both fixed points
  approach one another, i.e., gravity has a destabilizing effect on
  the pure matter system. Finally, at a critical
  gravitational coupling strength $g_\text{crit.} \approx 6$ the two 
fixed points have been driven to a collision by metric
fluctuations and no real fixed point remains. 

It is thus crucial to understand the system of induced interactions in
more detail, to clarify whether asymptotically safe gravity is
compatible with the existence of a Yukawa sector in Nature. In this
spirit, we provide a first step by analyzing the full system in the
next section.

  \begin{figure*}[t]
		\begin{center}
			\includegraphics[width=0.34\linewidth]{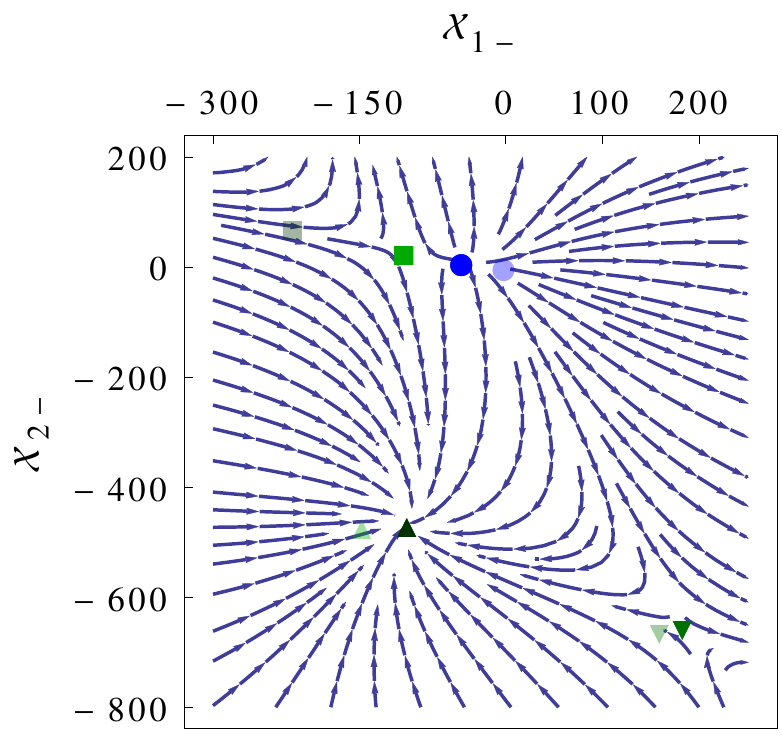}\hfill
			\includegraphics[width=0.315\textwidth]{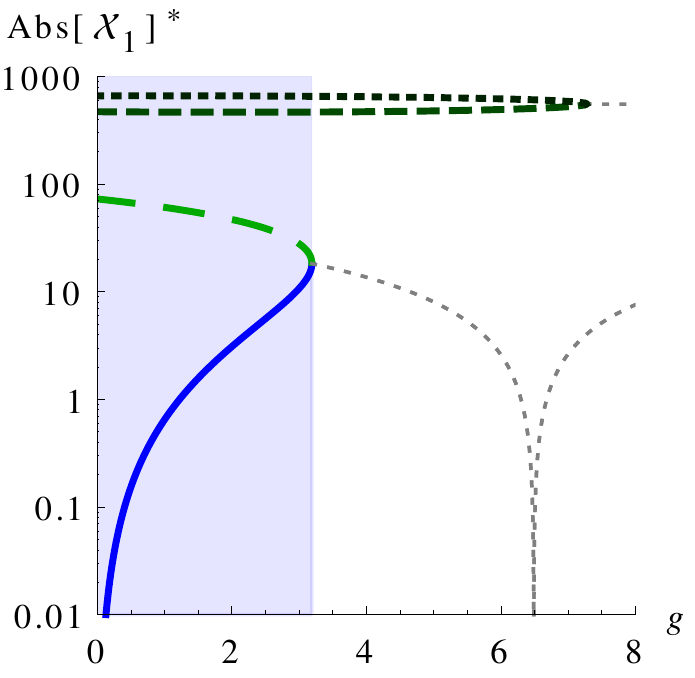}\hfill
			\includegraphics[width=0.315\textwidth]{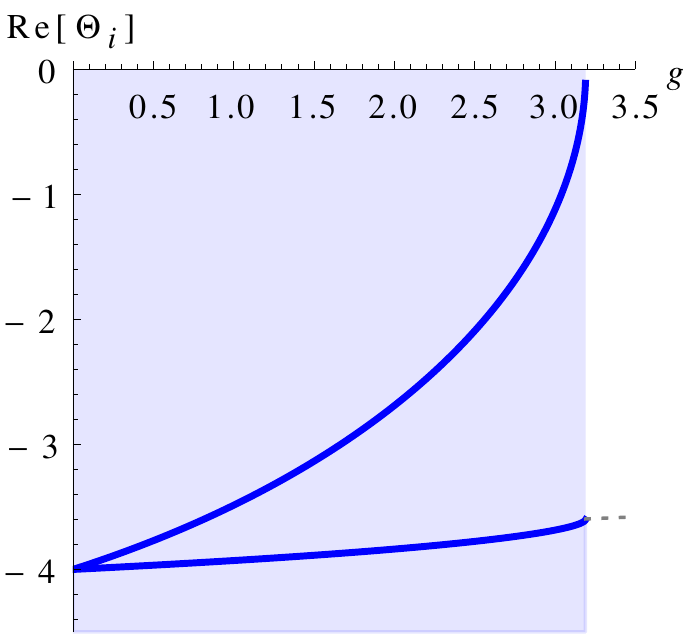}
		\end{center}
                \caption{Left-hand panel: Flow lines towards
                  the UV for $g=3$ in
                  $\mathcal{X}_{1-}\,\mathcal{X}_{2-}$-space.  The
                  four fixed points sGFP, NGFP1, NGFP2 and NGFP3 are
                  marked by a circle, square, up-triangle and
                  down-triangle respectively.  Lighter shaded points
                  mark the locations of the fixed points at $g=0$.
                  \\
                  Middle panel: Change of fixed-point coordinates
                  $\mathcal{X}_{1-}^\star$ of the sGFP (solid blue),
                  NGFP1 (dashed green), NGFP2 and NGFP3 (dotted dark
                  green) with growing $g$. Real parts of complex
                  fixed-point values are shown in light gray.
                  \\
                  Right-hand panel: Real parts of the critical
                  exponents of the shifted Gau\ss ian fixed point
                  (sGFP) for growing $g$. The blue-shaded area marks
                  the region where all fixed point coordinates are
                  real. The fixed-point collision at $g\approx 3.2$
                  (cf. right-hand panel) is accompanied by a vanishing
                  critical exponent. }
	\label{fig:x3mx4mEvol}
\end{figure*}

\subsection{Induced fixed-point interactions: the full system}
\label{sec:fully_disentangled_truncation}
\subsubsection{Shifted Gau\ss{}ian fixed point and fixed-point
  annihilations} 

After the instructive, simplified analysis in the reduced system in
the previous Section, we now take into account both induced
two-fermion--two-scalar interactions as put down in
\eq{eq:YukawaXre}. To disentangle all tensor structures in the
$\mathcal{X}$-sector we now use a projection on the two couplings
separately,
cf.~Eqs.~\eqref{eq:projX3m},\eqref{eq:projX4m},\eqref{eq:projX3p} and
\eqref{eq:projX4p}. After projecting out the terms with three momenta,
we choose a fully symmetric momentum configuration,
cf.~App.~\ref{app:proj}.  The resulting beta-functions are presented
in App.~\ref{app:Xbetas}.

For vanishing Newton coupling, $g=0$, the two-dimensional matter
system of non-vanishing couplings exhibits four fixed points. In
addition to the Gau\ss ian fixed point three interacting
fixed points are purely real and thus potentially physical. Values of
all real fixed points and critical exponents for specific
  choices of gravitational couplings can be found in
Tab.~\ref{tab:FP-X3m-X4m}.

\begin{table}[h]
\begin{tabular}{c|c|c|c|c|c}
	$g$&$\mathcal{X}_{1-}^*$&$\mathcal{X}_{2-}^*$&$\theta_1$ & $\theta_2$\\
	\hline \hline
	0 & 0 & 0 & -4 &-4 & GFP\\
	& -217.9 & -72.7 & 4&-3.55& $\rm NGFP1$\\
	& -145.9 &-469.4 & 4.59& 4 & $\rm NGFP2$\\
	& 159.7 & -660.2 & 4& -5.38& $\rm NGFP3$\\ 
	\hline \hline
	2&	-12.74 & 3.08 & -2.69& -3.84& sGFP\\
	& -157.7 & 46.9 & 2.68& -3.49& $\rm NGFP1$\\
	& -118.5 &-464.8 & 4.10 + 0.49 i & 4.10 - 0.49 i & $\rm NGFP2$\\
	& 178.6 & -658.4 & 3.92& -5.25 &$\rm NGFP3$\\ 
	\hline \hline
	4 & -76.8 & -469.7 & 3.67 - 0.95 i &-3.67 + 0.95 i & NGFP2\\
	& 183.8 & -647.0 & 3.78 & -4.64& $\rm NGFP3$\\
	\hline \hline
\end{tabular}
\caption{\label{tab:FP-X3m-X4m} Fixed-point values and critical exponents 
  in the $\mathcal{X}_{1-}$-$\mathcal{X}_{2-}$-truncation for $g=0$, $g=2$ and $g=4$, for vanishing graviton mass parameter and vanishing anomalous dimensions.}
\end{table}
  
In a first step, we keep $\mu_h=\eta_h=\eta_\phi=\eta_\psi=0$ and only
vary the gravitational interactions with $g$. In this process gravity  
shifts all four real fixed points (sGFP, NGFP1/2/3) of the
pure matter system, cf.~Fig.~\ref{fig:x3mx4mEvol} and Tab.~\ref{tab:FP-X3m-X4m}. 

The shifted Gau\ss ian fixed point (sGFP) remains UV repulsive in
both directions but becomes non-Gau\ss ian when gravity is
switched on, cf.~Fig.~\ref{fig:x3mx4mEvol}.  The gravitational
contribution to the scaling dimensionality of $\mathcal{X}_{1/2-}$ is positive for both of these
couplings. Accordingly, they are shifted towards relevance. At $g \approx 3.2$, the shifted Gau\ss ian fixed point collides with
one of the other interacting fixed points, and moves off into the
complex plane. As expected, this fixed-point collision is accompanied
by a change in sign in one of the critical exponents,
cf.~Fig.~\ref{fig:x3mx4mEvol}, right panel. Thus gravity
might not only deform the universality class of an asymptotically free
matter fixed point, but might even completely destroy it. In a partial
range of values for $g$ where the shifted Gau\ss ian fixed point does
not exist, two other interacting fixed points remain
at real coordinates.  These provide possible universality classes for
the matter system coupled to gravity. As the presence of additional
interacting fixed points is a difference to the simpler truncation
analyzed in Sec.~\ref{simpletruncation}, it is not clear whether these
are truncation artifacts.

At $g \approx 7.3$ also the other two interacting fixed points
  collide and for effective metric fluctuation-strengths stronger than
  that no fixed point can be found at all. It is unclear whether the
  vanishing of all possible fixed points is a truncation artifact but
  if this persists in higher-order truncations it implies that the
  strength of quantum-gravity interactions must not exceed a critical
  value, as otherwise the existence of a Yukawa sector for matter is
  excluded. 

Our results further exemplify, that the regime of very
strong gravitational coupling could correspond to a setting, where the
UV completion for the Standard Model coupled to gravity might be one
which is fully non-Gau\ss ian, and which is actually \emph{not} a
quantum-gravity deformation of an asymptotically free fixed point.

\subsubsection{Allowed gravitational parameter space}

The effective gravitational interaction strength is related to the
combination $g/(1+\mu_h)$, and thus grows with increasing  $g\geqslant0$ and
decreasing $\mu_h\geqslant-1$. Thus at a very large mass parameter, the shifted
Gau\ss ian fixed point is nearly Gau\ss ian. As a function of
decreasing mass parameter, it then undergoes the same type of
collisions as it does as a function of increasing $g$. We thus
conclude that the shifted Gau\ss ian fixed point does not exist in all
regions of the $g\, \mu_h$-plane
within our truncation, cf.~Fig.~\ref{fig:xFPexistencePlane}. 
If we assume that a phenomenologically viable
fixed point has to be the shifted Gau\ss{}ian one, in order to
smoothly connect to the perturbative behavior of the Standard Model in
the vicinity of the Planck scale, the gray region in
Fig.~\ref{fig:xFPexistencePlane} is exluded from the viable
gravitational parameter space in our truncation.

Comparing these results to the simpler truncation in
Section~\ref{simpletruncation}, we observe that the region in which
the shifted Gau\ss ian fixed point is complex remains quantitatively
similar and persists. Whether this trend also persists under further
extensions of the truncation is an important question for future
studies.

Matter anomalous dimensions shift the areas of existence of the
shifted GFP as they shift the critical exponents and therefore also
the locations where these cross zero.  Interestingly, the fixed point
values of the non-interacting matter-gravity sector
\cite{Meibohm:2015twa} fall into a regime of gravitational strength
where the shifted GFP of the $\mathcal{X}$-sector exists
(cf.~Fig.~\ref{fig:xFPexistencePlane}), with 
\bea
	\mathcal{X}_{1-}^\ast &=& -10.3\;,\quad
	\mathcal{X}_{2-}^\ast = 2.40\;,\nonumber\\
	\theta_{1}&=& -4.33, \quad \theta_2 = -3.23.
\eea
Note that these values will be
modified in a combination of the study in \cite{Meibohm:2015twa} with
our truncation, as the $\mathcal{X}$-interactions alter the matter
anomalous dimensions. As in Sec.~\ref{sec:stability}, large negative
matter anomalous dimensions can qualitatively alter our results. If,
e.g., $\eta_{\psi} + \eta_{\phi} \lesssim-2.5$ the shifted Gau\ss{}ian
fixed point disappears. It is therefore crucial to quantify shifts in
the matter anomalous dimensions due to the induced matter
interactions. If such large shifts are indeed observed, truncations
should include these interactions, even though they are of higher
order according to our ordering principle of canonical dimensionality.

Combining the two aspects of the Yukawa sector that we have analyzed,
we observe that the $\mathcal{X}$-couplings cannot couple directly into the
flow of the Yukawa coupling, due to their derivative structure. Thus,
a truncation including $y, \mathcal{X}_{1-}$ and $\mathcal{X}_{2-}$ still features the Gau\ss
ian fixed point for $y$.  This fixed point can be combined with any of
the fixed points in the $\mathcal{X}$-sector, which remain unaltered
by the inclusion of $y$. On the other hand, a more extended
truncation, including momentum-dependent Yukawa couplings, as well as
two-fermion--two-scalar couplings with a lower power of momenta could
feature contributions of these new couplings to both $\beta_y$ as well
as $\beta_{\mathcal{X}_{1/2-}}$.

\section{Conclusions and Outlook}
\label{sec:conclusions}

As a step towards a unification of the Standard Model with quantum
gravity within the paradigm of asymptotic safety, we have performed an
extensive analysis of a Yukawa system. Our model consists of a Dirac
fermion and a scalar, coupled to asymptotically safe quantum
gravity. The fixed-point structure of our model is determined by two
major quantum-gravity effects on the matter sector:
\begin{enumerate}
\item Quantum-gravity fluctuations generate a correction to the
  quantum scaling of matter operators.
\item Quantum-gravity fluctuations induce non-zero matter self
  interactions, triggering a departure from asymptotic freedom in the
  matter sector.
\end{enumerate}

The first property is of particular interest in a scenario where no
additional matter degrees of freedom exist up to the Planck
scale. There, a direct connection can be established between
low-energy values of the Standard Model couplings, and the properties
of a joint matter-gravity fixed point. In particular, the low-energy
values of irrelevant couplings can be predicted, thus potentially
enabling observational tests of the quantum-gravity regime. Our
studies suggest that the Yukawa coupling could be used to bridge the
gap between experimentally accessible scales and the Planck scale: In
the UV, the Yukawa coupling in our toy model features a UV-repulsive
fixed point at zero for most of the gravitational parameter space.
This implies that it must already be very close to zero at the Planck
scale. This condition is actually realized in Nature for the Yukawa
couplings of the lighter leptons of the Standard Model. However the
top Yukawa coupling is sizable at the Planck scale, which would
probably prevent it from reaching a fixed point at zero in the far UV.

\begin{figure}[t]
  \includegraphics[width=\linewidth]{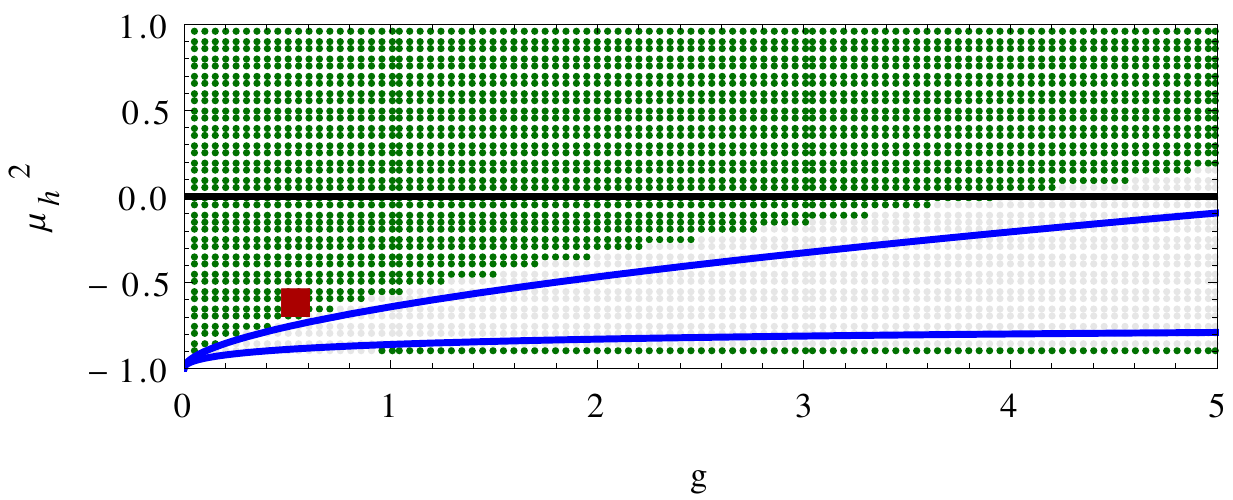}
  \caption{ Existence of a fully attractive fixed point corresponding
    to the shifted Gau\ss ian fixed point in the fully disentangled
    $\mathcal{X}_i$-truncation for different values of the
    gravitational parameters $(g,\mu_h)$ and vanishing anomalous
    dimensions. Darker green dots in the $g,\mu_h$ plane mark
    existence regions. For the gravitational fixed point values
    $g_\ast \approx 0.55$ and $\mu_{h,\ast} \approx -0.58$ (red dot)
    from \citep{Meibohm:2015twa} an interacting matter fixed point
    with two irrelevant directions exists. The depicted fixed-point
    values have been obtained for non-vanishing anomalous dimensions
    taken from \cite{Meibohm:2015twa}.  Inbetween the two thick blue
    lines, no shifted GFP exists in the simplest truncation,
    cf.~Fig.~\ref{fig:betaX}. We observe qualitative agreement between
    the different approximations.  }
	\label{fig:xFPexistencePlane}
\end{figure}

Within our toy model, there is a small region of parameter space for
which the Yukawa coupling becomes relevant. Simultaneously, an
interacting fixed point at which it is irrelevant is generated. If
extended truncations feature a similar fixed point, a scenario is
conceivable, in which the observed value of the top-Yukawa coupling is
in fact a consequence of asymptotic safety.

For the Yukawa sector, the second major quantum-gravity effect is
manifest in newly generated, momentum-dependent interactions. The
lowest order gravity-induced interactions between fermions and scalars
are two-fermion--two-scalar interactions with couplings
$\mathcal{X}_i$. At vanishing gravitational coupling, the system
features a non-interacting, Gau\ss{}ian fixed point.  Then, switching on
quantum gravity fluctuations, the Gau\ss ian matter fixed point is
shifted, becoming an interacting fixed point which we call shifted
Gau\ss ian fixed point. In an extended truncation, we thus find only
interacting fixed points for the $\mathcal{X}_i$, while the Yukawa
coupling $y$ is zero at that fixed point. Thus, the only viable
matter-gravity fixed point appears to be one which is asymptotically
safe, rather than asymptotically free, at least for a subsector of
matter self-interactions. Unlike other interacting fixed points of the
matter system, the shifted Gau\ss ian fixed point shows scaling
exponents similar to the canonical ones, at least in the weak-gravity
regime.

We have demonstrated that the model is dominated by an intriguing
interplay of several fixed points at large values of the gravitational
coupling. At a critical value of the gravitational coupling strength,
one of these interacting fixed points collides with the shifted Gau\ss
ian fixed point. Subsequently, they disappear into the complex
plane. Thus, the shifted Gau\ss ian fixed point ceases to exist for
larger values of the effective gravitational coupling.  In fact,
gravitational fixed-point values from recent studies lie within the
regime allowing a shifted Gau\ss ian fixed point,
cf.~Fig.~\ref{fig:xFPexistencePlane} which summarizes our main
results.  Crucially, extending our truncation leads to quantitatively
similar results on the allowed gravitational parameter space,
cf.~Fig.~\ref{fig:xFPexistencePlane}.

Our results highlight that even though the canonical dimensionality of
the momentum-dependent matter interactions is irrelevant, they might
nevertheless play a pivotal role in the dynamics of an asymptotically
safe matter-gravity system.  We conclude that quantum gravity could
have a significant impact on the properties of the matter sector in
the UV, and might even require a UV completion that deviates
considerably from a shifted Gau\ss ian fixed point and the associated
near-canonical power counting. In order for this scenario to be
realized, gravitational couplings must exceed a critical strength,
which lies beyond that observed in the literature, see, e.g.,
\cite{Becker:2014qya,Meibohm:2016mkp,Dona:2013qba}.  Thus, truncations
that neglect gravity-induced matter self interactions, and therefore
discover a free fixed point in the matter sector might potentially not
capture the properties of a matter-gravity fixed point at a
qualitative level: If the joint matter-gravity fixed point is not the
shifted Gau\ss ian one, then the scaling exponents in the matter
sector will be rather different from those at a Gau\ss ian matter
fixed point. 

As one of our main results, we find that a shifted
Gau\ss{}ian fixed point with irrelevant fermion-scalar interactions
and an irrelevant Yukawa coupling exists at gravitational fixed-point
values from recent studies.  

\subsection{Outlook}

Our study also clearly indicates the need for extended truncations in
the matter sector. In the future, we will extend our current study to
include further matter interaction terms that have an impact on the
scaling dimension of the Yukawa coupling and on the induced
matter-interactions.  A major missing piece of this puzzle are
four-fermion interactions, which have been shown to feature only
interacting fixed points under the effect of asymptotically safe
quantum gravity \cite{Eichhorn:2011pc,Meibohm:2016mkp}. They feed back
into the beta-functions in our system.  Further, momentum-dependent
self-interactions of scalars and fermion-scalar interactions will
alter the value of the matter anomalous dimensions. As has been
discussed in Section~\ref{sec:quartic}, including dynamical anomalous
dimensions is a critical step towards a quantitative control of the
system.

It should be stressed that within asymptotically safe quantum gravity,
the Einstein-Hilbert action is not the complete fixed-point action for
gravity. Just as quantum fluctuations induce further matter
interactions, higher-order terms, such as, e.g., curvature-squared
terms, are also present. In particular, a subset of these corresponds
to relevant couplings at a pure-gravity fixed point. Testing the
impact of these terms on the matter sector is an important bit of the
puzzle that we hope to come back to in the future.

On the phenomenological side, the quantum-gravity-generated
fermion-scalar interactions could be of interest in the context of
Higgs portals to fermionic dark matter, see, e.g.,
\cite{Beniwal:2015sdl}: Asymptotically safe quantum gravity, according
to our studies, contains a generic mechanism to couple dark sectors,
e.g., additional scalars and fermions, to a scalar such as the
Higgs. This coupling might therefore provide a basis for a possible
connection of asymptotic safety to dark matter phenomenology.\\[-2ex]


\emph{Acknowledgements}\\ We thank Masatoshi Yamada and Holger
  Gies for discussions on \cite{Oda:2015sma} and reflection
  positivity. We also thank Omar Zanusso and Manuel Reichert, Tobias
Henz for discussions as well as Anton Cyrol, Mario Mitter, Andreas
Rodigast and Tobias Denz for the (co-) development of the automated
algebraic tools used in this computation \cite{Mitter:2014wpa,
  Cyrol:inPrep, Denz:inPrep}. This work is supported by EMMI and by
ERC-AdG-290623. The work of A.E. is supported by an Imperial College
Junior Research Fellowship.
\newline\\

\appendix

\section{Background independence
}
\label{app:diffeo}

The classical action of the present matter-gravity theory is given by
the first three lines in \eq{eq:truncation} with $Z_\Phi=1$, and the
Einstein-Hilbert action, \eq{eq:EHaction}. This action is a function
of a single metric field $g$ and consequently, does not depend on the
choice of the background metric $\bar g$.  This statement generalizes
to expectation values of diffeomorphism invariant operators for any
quantized matter-gravity theory, based on a microscopic action that
depends on a single metric.  In turn, for diffeomorphism-variant
correlation functions such as those of the metric fluctuations $h$,
diffeomorphism invariance and background independence translate into
non-trivial Slavnov-Taylor identities (STIs) and Nielsen identities
(or split Ward identities), that depend on the background metric. This
leads to the counter-intuitive situation, that any
diffeomorphism-invariant and background-independent approximation to
the fluctuation dynamics breaks the underlying diffeomorphism
invariance and background independence of the theory. For further
discussions see
\cite{Christiansen:2014raa,Christiansen:2015rva,Meibohm:2015twa}.

In the present setting with an additional regulator term
(cf. App.~\ref{app:reg} for a discussion of the employed regulators),
the STIs turn into mSTIs and modified Nielsen identities (split Ward
identities), see, e.g., 
\cite{Reuter:1993kw,Litim:1998nf,Litim:2002ce,Litim:2002hj,Pawlowski:2003sk,%
Pawlowski:2005xe,Manrique:2009uh,Manrique:2010mq,Manrique:2010am,%
Folkerts:2011jz,Bridle:2013sra,Dietz:2015owa,Labus:2016lkh}. 
Importantly, regulator modifications scale with the cutoff $k$ and are
hence potentially dominant, in particular for the UV-relevant
parameters. This makes the distinction between background and
fluctuation quantities crucial. The flow of both, background and
fluctuation correlations is generated by closed flow equations of the
fluctuation propagator and vertices.  In turn, only the background
field effective action $\Gamma_{k=0}[ g;
\tilde\Phi=\Phi=(0,0,0,\psi,\bar\psi,\phi)]$ is diffeomorphism
invariant, background-independent and relates directly to S-matrix
elements of the theory.  By this reasoning -- in order to compute
diffeomorphism invariant and background-independent observables -- we
first have to solve the dynamical, closed system of fluctuation
correlation functions $\Gamma^{(n)}$ and use the latter in the flow of
$\Gamma_{k=0}[ g; \tilde \Phi]$.

\section{Expansion scheme}\label{app:expansion}

The vertex expansion reads 
\begin{align}\label{eq:VEXP}
  \Gamma_k[\bar{g},\Phi] =
  \sum_n\frac{1}{n!}\Gamma^{(n)}[\bar{g},0]\Phi^n \;,
      \end{align}
      and requires the choice of a specific metric and matter background 
      $(\bar g,\bar \Phi)$ as an expansion point. We
      choose $\bar\Phi=0$, which is a saddle point or minimum of the
      matter part of the effective action. A good physical choice for
      the metric background is a solution of the gravity equation of
      motion, which however complicates the computations considerably. 
Instead we choose a flat background, 
\begin{align}\label{eq:flat}
 \bar{g}_{\mu \nu} = \delta_{\mu \nu}
\end{align}
with a flat Euclidean metric. 
We emphasize that it is not
necessary to choose a solution of the equations of motion, $\bar
g_{\text{\tiny { EoM}}}, \bar \Phi_{\text{\tiny { EoM}}}$ as the
expansion point. It is expected, though, that such a choice
$\bar g_{\text{\tiny { EoM}}}, \bar \Phi_{\text{\tiny { EoM}}}$ leads
to a more rapid convergence of the vertex expansion.
\\

In the present work we use the lowest order of the vertex expansion in
the matter sector to access low-order $n$-point functions.  The
related effective matter-gravity action underlying our results in
Sec.~\ref{sec:gauge} is summarized in the first three lines of
\eq{eq:truncation} and amounts to $\Gamma_{k,\rm ho}=0$. This
approximation also features an additional derivative expansion, as no
higher momentum-dependences and additional tensor structures are
considered.  In its lowest order this leaves us with the
scale-dependent Yukawa coupling $y$.  Additionally, wave-function
renormalizations $Z_{\phi/\psi}$ are included. At the next order in
our expansion scheme, these would be upgraded to field-dependent
wave-function renormalizations.  In this work, we neglect additional
terms in the purely scalar sector which are canonically relevant/
marginal, as they do not directly impact $\theta_y$, which is the main
physics question we focus on.

In Sec.~\ref{sec:quartic}, we consider two-fermion--two-scalar
interactions, induced by gravity. These interactions give rise to
additional terms in the effective action, $\Gamma_{k,\rm induced}$
that are part of the higher order terms in $\Gamma_{k,\rm ho}$. From
the class of all gravity-induced interactions, these are the
lowest-order terms in a combined derivative- and vertex expansion,
i.e., they are those with the least irrelevant canonical
dimensionality. Following a canonical counting, lower-order
scalar-fermion interactions exist; however, these are not induced by
gravity fluctuations directly. On the other hand, once gravity
fluctuations switch on the $\mathcal{X}_i$ interactions, these can
induce further matter interactions, e.g., through tadpole
diagrams. This backcoupling of quantum-gravity induced interactions
will be the subject of a future work.

Finally, by using a metric split, cf.\eq{eq:gbargh}, in the
matter part of the effective action (cf.~\eq{eq:truncation}) we
have identified all matter couplings to gravity with a single Newton
coupling $\sqrt{G}$. Typically, such a universality holds for the
first two non-vanishing perturbative coefficients of the
$\beta$-function of a dimensionless coupling in a mass-independent
renormalization scheme.  None of the above hold in the current
case. Instead, it is expected that all the gravity couplings run
differently due to the missing universality as well as due to the
regulator-induced modifications of the mSTIs.  The respective
modifications are either of higher order in the gravity fluctuation
$h$, as well as higher order in derivatives.

Our computations rely on automated algebraic manipulation tools. For
setting up the truncation as well as deriving the vertices and
diagrams we use the Mathematica package TARDIS
\cite{Denz:inPrep}. When calculating the traces of these diagrams we
rely on the FormTracer package \cite{Mitter:2014wpa, Cyrol:inPrep}.

\section{Gravity expansion}\label{app:gravity}

The above Section~\ref{app:expansion} summarizes our expansion scheme
in the matter sector of the theory.  For the fixed point analysis we
also need the metric propagator as well as the Newton coupling in a
flat background. Given the expansion in powers of the fluctuating
graviton $h$, this is naturally augmented with results in the same
expansion scheme for the gravity correlations put forward in
\cite{Christiansen:2012rx,Christiansen:2014raa,Christiansen:2015rva,%
  Meibohm:2015twa,Meibohm:2016mkp}, or with mixed approaches
\cite{Codello:2013fpa,Dona:2013qba,Dona:2015tnf}. Equivalently, the
same information can be encoded in an approach featuring the
background metric as well as the full metric,
\cite{Manrique:2009uh,Manrique:2010mq,Manrique:2010am,Becker:2014qya}.
Note that our results for the matter sector can be combined with
different approximation schemes in the gravity sector, as they simply
use the gravitational couplings $g$, $\mu_h$ and $\eta_h$ as input
parameters.

  We adopt the linear split 
  (cf.~\eq{eq:gbargh})
  in the Einstein-Hilbert action (cf.~\eq{eq:EHaction}) in the 
  spirit of the
  present combined vertex and derivative expansion, and drop higher
  order terms. 
  Full consistency of the present expansion scheme as a
  derivative expansion then also requires $\alpha\to Z_\alpha \alpha,
  \, \beta\to Z_\beta \beta$ in the gauge fixing term
  \eq{eq:gauge-condition}. With these cutoff dependences the quadratic
  part of the pure gravity effective action reads
\begin{align} \nonumber &\, 
\Gamma_{k\, \rm grav}^{(h,h)\mu\nu\rho\sigma}h_{\mu\nu}h_{\rho\sigma}= 
\frac{Z_h}{64\pi}
  \int \frac{d^4p}{(2 \pi)^4}\,h_{\mu\nu}(p) h_{\rho\sigma}(-p)
  \\[2ex] \nonumber &\, \times\Biggr[ \left(\mu_h
    k^2+p^2\right)g^{\mu(\rho}g^{\sigma)\nu} -\left(\mu_h
    k^2+2p^2\right)g^{\mu\nu}g^{\rho\sigma} \\[2ex] \nonumber
  &\,+2\left(g^{\mu\nu}p^\rho p^\sigma + g^{\rho\sigma}p^\mu
    p^\nu\right) -g^{\mu(\rho}p^{\sigma)}p^\nu -
  g^{\nu(\rho}p^{\sigma)}p^\mu \\[2ex] \nonumber
  &\,-\0{1}{Z_\alpha}\Biggl(\frac{1+Z_\beta
    \beta}{\alpha}\left(g^{\mu\nu}p^\rho p^\sigma +
    g^{\rho\sigma}p^\mu p^\nu\right) \\[2ex] &\,+\frac{1}{
    \alpha}\left(g^{\mu(\rho}p^{\sigma)}p^\nu -
    g^{\nu(\rho}p^{\sigma)}p^\mu\right) +\frac{(1+Z_\beta
    \beta)^2}{4\, \alpha}p^2g^{\mu\nu}g^{\rho\sigma} \Biggr)\Biggr]\,.
\label{eq:hh-app}
\end{align}
The present approximation is based on a common approximation on the
mSTI in gauge theories, leading to a diffeomorphism-invariant
representation with multiplicative wave function and coupling
renormalizations. In this work it is induced by the truncation in
(cf.~\eq{eq:truncation}) with linear split
(cf.~\eq{eq:gbargh}) and by using diffeomorphism-invariant terms
in the effective action before substituting \eq{eq:gbargh}. This
leads to a uniform wave function renormalization $Z_h$ for all York
decomposed components of the graviton propagator (cf. line 2 \& 3 in
\eqref{eq:hh-app}), except for the gauge fixing terms (cf. last
two lines in \eqref{eq:hh-app}).  The York decomposition reads
\begin{align}
	\label{eq:york}
        h_{\mu\nu}=h^\text{TT}_{\mu\nu} 
        + 2\bar{D}_{(\mu}v_{\nu)}
        + \left( 2\bar{D}_{(\mu}\bar{D}_{\nu)}-\frac{1}{2}\bar{g}_{\mu\nu}\bar{D}^2\right)\sigma
        +\frac{1}{4}\bar{g}_{\mu\nu} h,
\end{align}
with $\bar{D}^{\mu}h_{\mu \nu}^{\rm TT}=0$, $h^{\mu\, \, \, \rm
  TT}_{\mu}=0$, $h^{\mu}_{\mu} =h$ and $\bar{D}^{\nu}v_{\nu}=0$.  In
the gravity results used here $Z_h$ is obtained from the traceless
transverse part of the inverse propagator, $Z_h= Z_{h,\rm TT}$.

Moreover, the related approximation of the mSTI leads to  
\begin{align}\label{eq:ZaZb} 
Z_\alpha=Z_h\,,\qquad \qquad Z_\beta=1\,,
\end{align} 
for the longitudinal directions singled out by the gauge condition
in \eq{eq:gauge-condition}. For the sake of simplicity we have used the additional approximation $Z_\alpha=1$ for the results shown in this
work. 
We have checked that the results do not differ
significantly from the consistent choice \eq{eq:ZaZb}.

The respective results for gravity in the literature are mostly
obtained in very few gauge choices, singled out by conceptual or
technical reasons. For instance, the harmonic gauges, $\beta=1$, or
the Feynman gauge, $\alpha = \beta = 1$, lead to considerable
technical simplifications. However, none of these gauges persists
during the flow for $\alpha\neq 0$ beyond the current, classical,
approximation to the mSTIs. In turn, $\alpha=0$ is a fixed point of
the FRG flow, as the flow of $1/\alpha$ is finite, see
\cite{Litim:1998nf}. In terms of convergence of the current
expansion in the gauge fixing sector of the theory this singles out
the line $(\alpha,\beta)=(0,\beta)$ as the most stable
one.
\\

The choice of $\beta$ also rotates the physical mode in the scalar
sector from pure trace-mode ($\beta \rightarrow 0$) into pure
$\sigma$-mode (see Eq. (21) in \cite{Gies:2015tca}). Inverting the
scalar projection of the 2-point function onto the scalar York-sector
$\Gamma^{(2)}_{k,\text{grav},\;(\sigma h)}[\Phi]$, the corresponding
part of the metric propagator is given by
\begin{align}
\label{eq:h-prop}
\Gamma^{(2)-1}_{k,\text{grav},\;(\sigma h)}
&=
\frac{64 \pi  G}{4 \alpha  \mu _h^2
+2 p^2 \left(2 \alpha -\beta ^2+3\right) \mu _h
+(\beta -3)^2 p^4}\times\notag\\[2ex]
\times &\begin{pmatrix}
	p^2 (\alpha -3)-2 \alpha  
	{\mu_h}& 
	\sqrt{3} p^2 (\alpha -\beta ) \\
 	\sqrt{3} p^2 (\alpha -\beta ) & 
 	\left(3 \alpha -\beta ^2\right) p^2+2 \alpha 
	{\mu_h}
\end{pmatrix}\,.
\end{align}
This equation already shows that for $\beta=3$ the gauge-fixing is
incomplete.  In terms of RG-flows also the gauge-parameters should in
fact be regarded as running 
quantities (directions in theory space). This viewpoint favors
$\alpha^\star=0$ (Landau-limit) as a UV-fixed point of the RG-flow
\citep{Litim:1998nf}. 
\\

Finally concerning the gauge-dependence of flows in the matter sector, e.g., the Yukawa coupling, it is interesting to observe that they only enter via the metric propagator or its 
regularized
version
\begin{align}
	&\left(\Gamma_k^{(2)}[\bar{g}_{\mu \nu}, \Phi]+ R_k\right)^{-1}
	\partial_t R_k
	\left(\Gamma_k^{(2)}[\bar{g}_{\mu \nu}, \Phi]+ R_k\right)^{-1}
	\nonumber\\[2ex]
	=\;
	&\frac{1}{\left((\beta -3)^2+2 \mu _h \left(2 \alpha -\beta ^2+2 \alpha  \mu _h+3\right)\right){}^2}\times
	\nonumber\\[2ex]
	&\times
	\frac{1}{\left(\mu _h+1\right){}^2} 
	\frac{1}{\left(\alpha  \mu _h+1\right){}^2}
	\;\times\;\textit{analytic}\;.
	\label{eq:gravPoleStructure}
\end{align}
Since in all diagrams in Fig.~\ref{yukawadiags} there is precisely one -possibly regulated- metric propagator, the gauge-pole structure of the Yukawa $\beta$-function is at most given by the above regulated propagator poles.\\
Indeed all gauge poles as well as poles in $\mu_h$ are lifted when dividing \eqref{eq:gauge-beta-yukawa} by the pole structure given in \eqref{eq:gravPoleStructure}.

\section{Optimized regulators} \label{app:reg} 

For the explicit computations in the flat background we employ
the Litim or flat cutoff \cite{Litim:2000ci,Litim:2001up} in momentum
space for the matter degrees of freedom,
\begin{align} \label{eq:regs} 
  R_{k,\phi}(p)= Z_{\phi}\, p^2 r_{\phi}(x)\,, \qquad R_{k,\psi}(p)=
  Z_{\psi}\, \slashed{p}\, r_{\psi}(x)
\end{align}
with $x=p^2/k^2$. The shape functions for bosons and fermions take the
form
\begin{align} \label{eq:shape} 
r_{\phi}(x) = (x-1)\theta(1-x)\,,\qquad
  r_{\phi}(x) = (\sqrt{x}-1)\theta(1-x)\,,
\end{align}
The graviton propagator that enters the diagrams for the matter wave
function renormalizations and the Yukawa coupling is represented in
the York decomposition. The regulator used for all modes in the York
decomposition (cf.~\eqref{eq:york}) is the bosonic one in
\eqref{eq:regs}, where $Z_\phi\, p^2$ is substituted by the full
kinetic factor that can be read-off from \eqref{eq:hh-app}.  A
comparison of different, also smooth, regulators for pure gravity
computations in a flat background has been put forward in
\cite{Christiansen:2012rx}. The regulator dependences of the results
have been found to be small.

We close this section with a remark on the use of flat regulators in a
derivative expansion. The flat regulator has been singled out by
optimization theory as the the optimal one within the lowest order of
the derivative expansion, see \cite{Litim:2000ci,Pawlowski:2005xe}. To
higher orders in the derivative expansion this does not hold because
of the non-analyticity of the regulator. Note however, that smooth,
analytic versions of the flat regulator satisfy the optimization
criterion for higher order of the derivative expansion, see
\cite{Pawlowski:2005xe}. The latter property is potentially
important for the convergence of our approximation, when we take
higher order gravitationally induced terms into account, see
Sec.~\ref{sec:quartic}. It turns out that in our projection scheme all
derivatives with respect to external momentum hit the
momentum-dependences of the vertices. In other words, in terms of
optimization we are still in the same situation as for the lowest
order of the derivative expansion, and the regulators satisfy the
optimization criteria in \cite{Litim:2000ci,Pawlowski:2005xe}.

\section{Gauge-parameter dependence of the Yukawa $\beta$-function}
\label{app:gauge-beta-yukawa}
Here we present the full gauge-dependent $\beta_y$-function for $\mu_{\phi}=0$. All other parameters are kept free.
\begin{widetext}
\begin{align}
	\label{eq:gauge-beta-yukawa}
	\beta_y &= y \left(\eta _{\psi,0}+\frac{\eta
            _{\phi,0}}{2}\right)+\frac{y^3 \left(5-\eta _{\psi
            }\right)}{80 \pi ^2 \left(\mu _{\phi }+1\right)}+\frac{y^3
          \left(6-\eta _{\phi }\right)}{96 \pi ^2 \left(\mu _{\phi
            }+1\right)^2} \\[1ex]& +y g \Biggl[ +\frac{ \left(\eta _{\psi
            }-6\right) \left(\beta -2 \alpha \mu _h-3\right)}{5 \pi
          \left(6 \beta -2 \mu _h \left(2 \alpha \left(\mu
                _h+1\right)+3\right)+\beta ^2 \left(2 \mu
              _h-1\right)-9\right)} \notag\\[1ex]& -\frac{12 \left(\eta
            _h-7\right) \left((\beta -3)^3-4 \alpha \left(2 \alpha
              -\beta ^2+\beta \right) \mu _h^2-4 \alpha (\beta -3)^2
            \mu _h\right)}{35 \pi \left(-6 \beta +2 \mu _h \left(2
              \alpha \left(\mu _h+1\right)+3\right)+\beta ^2 \left(1-2
              \mu _h\right)+9\right)^2} \notag\\[1ex]& -\frac{\eta _h-8}{64
          \pi \left(\alpha \mu _h+1\right)^2 \left(-6 \beta +2 \mu _h
            \left(2 \alpha \left(\mu _h+1\right)+3\right)+\beta ^2
            \left(1-2 \mu _h\right)+9\right)^2}\times \notag\\&
        \quad\times[(1-\alpha ) (\beta -3)^4+4 \alpha ^3 (4 \alpha -3
        (\beta -2) \beta -7) \mu _h^4 +\mu _h^2+8 (\alpha -1) \alpha
        ^2 (\beta -3)^2 \mu _h^3+2 \alpha (\beta -3)^2 (-4 \alpha +3
        (\beta -2) \beta +7) \mu _h \notag\\& \quad+\alpha \left(-16
          \alpha ^2+\alpha (\beta (\beta ((\beta -12) \beta
          +78)-156)+121)-4 \left(\beta \left(\beta ^3-3 \beta
              -6\right)+12\right)\right)] \notag\\[1ex]& +\frac{ \left(\eta
            _{\psi }-7\right) \left((\alpha -1) (\beta -3)^2+\alpha (4
            \alpha -3 (\beta -2) \beta -7) \mu _h\right)}{56 \pi
          \left(\alpha \mu _h+1\right) \left(-6 \beta +2 \mu _h
            \left(2 \alpha \left(\mu _h+1\right)+3\right)+\beta ^2
            \left(1-2 \mu _h\right)+9\right)} \notag\\[1ex]& -\frac{
          \left(\eta _h-6\right)}{12 \pi \left(\mu _h+1\right)^2
          \left(\alpha \mu _h+1\right)^2 \left(-6 \beta +4 \alpha \mu
            _h^2+(4 \alpha +6) \mu _h+\beta ^2 \left(1-2 \mu
              _h\right)+9\right)^2}\times \notag\\[1ex]& \quad \times [96
        \alpha ^4 (\mu _h+1)^2 \mu _h^4 -4 \alpha ^3 (\mu _h+1) \mu
        _h^2 (\beta ^2 (2 \mu _h^3+18 \mu _h^2-15 \mu _h-1)+6\beta (4
        \mu _h^2+15 \mu _h+1)-3 (6 \mu _h^3+62 \mu _h^2+73 \mu _h+7))
        \notag\\& \quad +\alpha ^2 \mu _h (3 \beta ^4 \mu _h (6 \mu
        _h^2-8 \mu _h+1)+12 \beta ^3 \mu _h (\mu _h^2+12 \mu
        _h-4)+\beta ^2 (-64 \mu _h^4-412 \mu _h^3-240 \mu _h^2+410 \mu
        _h+32)) \notag\\& \quad +\alpha^2(-12 \beta (31 \mu _h^3+148
        \mu _h^2+148 \mu _h+16)+3 (64 \mu _h^4+614 \mu _h^3+1272 \mu
        _h^2+833 \mu _h+96)) \notag\\& \quad +\alpha (3 \beta ^4 (4
        \mu _h^4+16 \mu _h^3-19 \mu _h^2+1)+12 \beta ^3 (8 \mu _h^3+33
        \mu _h^2-8 \mu _h-3)-2 \beta ^2 (40 \mu _h^4+280 \mu _h^3+177
        \mu _h^2-416 \mu _h-83)) \notag\\& \quad +\alpha(-12 \beta (32
        \mu _h^3+173 \mu _h^2+200 \mu _h+29)+3 (44 \mu _h^4+480 \mu
        _h^3+1081 \mu _h^2+768 \mu _h+93)) \notag\\& \quad +3 (\beta
        ^4 (6 \mu _h^2-8 \mu _h+1)+4 \beta ^3 (\mu _h^2+12 \mu
        _h-4)+\beta ^2 (-44 \mu _h^2-48 \mu _h+86)-12 \beta (\mu
        _h^2+12 \mu _h+16)+3 (26 \mu _h^2+72 \mu _h+51))] \notag\\[1ex]&
        +\frac{2 \left(\eta _{\phi }-6\right) \mu _{\phi }
          \left(\alpha \left(2 \mu _h-1\right)+3\right)}{3 \pi
          \left(\mu _{\phi }+1\right)^2 \left(-6 \beta +4 \alpha \mu
            _h^2+(4 \alpha +6) \mu _h+\beta ^2 \left(1-2 \mu
              _h\right)+9\right)} \notag\\[1ex]& -\frac{12 \left(\eta
            _{\phi }-7\right) \mu _{\phi } \left(\beta -2 \alpha \mu
            _h-3\right)}{35 \pi \left(\mu _{\phi }+1\right)^2 \left(6
            \beta -2 \mu _h \left(2 \alpha \left(\mu
                _h+1\right)+3\right)+\beta ^2 \left(2 \mu
              _h-1\right)-9\right)} \notag\\[1ex]& +\frac{2 \left(\eta
            _h-6\right) \mu _{\phi } \left((3-\alpha ) (\beta -3)^2+4
            \alpha \left(3 \alpha -\beta ^2\right) \mu _h^2+4 \alpha
            (\beta -3)^2 \mu _h\right)}{3 \pi \left(\mu _{\phi
            }+1\right) \left(6 \beta -2 \mu _h \left(2 \alpha
              \left(\mu _h+1\right)+3\right)+\beta ^2 \left(2 \mu
              _h-1\right)-9\right)^2} \notag\\[1ex]& +\frac{12 \left(\eta
            _h-7\right) \mu _{\phi } \left((\beta -3)^3-4 \alpha
            \left(2 \alpha -\beta ^2+\beta \right) \mu _h^2-4 \alpha
            (\beta -3)^2 \mu _h\right)}{35 \pi \left(\mu _{\phi
            }+1\right) \left(6 \beta -2 \mu _h \left(2 \alpha
              \left(\mu _h+1\right)+3\right)+\beta ^2 \left(2 \mu
              _h-1\right)-9\right)^2}\Biggr] \notag
\end{align}
\end{widetext}

\section{Explicit form of matter-metric vertices from covariant kinetic terms}
\label{app:matter-vertices}
To explain the derivative structure of the quartic interactions we
present the matter vertices as obtained when taking appropriate
variations of the covariant kinetic terms. Note that for fermions also
the variations of the spin-connection
\begin{equation}
	\delta\Gamma_\mu = -\frac{1}{8}D_\alpha\delta g_{\beta\mu}[\gamma^\alpha,\gamma^\beta]
\end{equation}
and the gamma-matrices
\begin{equation}
	\delta\gamma^\mu = \frac{1}{2}\delta g^{\mu\nu}\gamma_\nu
\end{equation}
are taken into account.  For the description of spinors in gravity we
use the spin-covariant derivative $\gamma^\mu\nabla_\mu
=\gamma^\mu(\partial_\mu + \Gamma_\mu)$, where $\Gamma^\mu$ denotes
the spin connection, defined using the spin-base invariant formalism
introduced in \cite{Gies:2013noa,Lippoldt:2015cea}.  Note that the
results for the variations from the spin-base invariant formalism
agree with those obtained in the standard vielbein formalism within a
symmetric O(4) gauge
\cite{vanNieuwenhuizen:1981uf,Woodard:1984sj}. This allows us to
rewrite vielbein fluctuations in terms of metric fluctuations.

\begin{align}
	\label{eq:vert-hff}\notag
	&\left[\frac{\delta}{\delta\psi(p_{\psi_1})}
	\frac{\delta}{\delta\bar\psi(p_{\psi_2})}
	\frac{\delta}{\delta h_{ \mu \nu}(p_{h_1})}	
	\Gamma_k\right]_{\Phi=0} \\[1ex]
	&=\frac{1}{8} \Bigr[
	-2g^{\mu\nu} \left(\slashed{p}_{\psi_1} +\slashed{p}_{\psi_2} \right)
	\notag\\[1ex]&
	+ \gamma^\mu p_{\psi_1}^\nu + \gamma^\nu p_{\psi_1}^\mu 
	+ \gamma^\mu p_{\psi_2}^\nu + 2\gamma^\nu p_{\psi_2}^\mu
	\Bigr]
\end{align}

\begin{align}
	\label{eq:vert-hhff}
	&\left[\frac{\delta}{\delta\psi(p_{\psi_1})}
	\frac{\delta}{\delta\bar\psi(p_{\psi_2})}
	\frac{\delta}{\delta h_{ \mu \nu}(p_{h_1})}
	\frac{\delta}{\delta h_{ \rho \sigma}(p_{h_2})}
	\Gamma_k\right]_{\Phi=0} 
	\notag\\[1ex]
	=\frac{1}{8} \Bigr[
	&\left(g^{\mu\sigma}g^{\rho\nu}+g^{\mu\rho}g^{\nu\sigma}-g^{\mu\nu}g^{\rho\sigma}\right)
	\left(\slashed{p}_{\psi_1}+\slashed{p}_{\psi_2}\right)
	\Bigr]
	\notag\\[1ex]
	+\frac{1}{64} \Bigr[&
	\left(p_{\psi_1}^\mu+p_{\psi_2}^\mu\right)
	\left(4g^{\rho\sigma}\gamma^{\nu}-3g^{\nu\rho}\gamma^\sigma-3g^{\nu\sigma}\gamma^\rho\right)
	\notag\\[1ex]&
	+\left(p_{\psi_1}^\nu+p_{\psi_2}^\nu\right) 
	\left(4g^{\rho\sigma}\gamma^{\mu}-3g^{\mu\rho}\gamma^\sigma-3g^{\mu\sigma}\gamma^\rho\right)
	\notag\\[1ex]&
	+\left(p_{\psi_1}^\rho+p_{\psi_2}^\rho\right)
	\left(4g^{\mu\nu}\gamma^{\sigma}-3g^{\mu\sigma}\gamma^\nu-3g^{\nu\sigma}\gamma^\mu\right)
	\notag\\[1ex]&
	+\left(p_{\psi_1}^\sigma+p_{\psi_2}^\sigma\right) 
	\left(4g^{\mu\nu}\gamma^{\rho}-3g^{\mu\rho}\gamma^\nu-3g^{\nu\rho}\gamma^\mu\right)
	\Bigr]
	\notag\\[1ex]
	+\frac{1}{32} \Bigr[&
	+g^{\mu\rho}g^{\nu\sigma}\left(\slashed{p}_{h_1}-\slashed{p}_{h_2}\right)
	\Bigr]
	\notag\\[1ex]
	+\frac{1}{64} \Bigr[&
	\left(\slashed{p}_{h_1}-\slashed{p}_{h_2}\right)
	\left(
		g^{\mu\sigma}\gamma^\nu\gamma^\rho
		-g^{\mu\rho}\gamma^\nu\gamma^\sigma
		\right.\notag\\[1ex]&\quad\quad\quad\quad\quad\quad\left.
		-g^{\nu\rho}\gamma^\mu\gamma^\sigma
		-g^{\nu\sigma}\gamma^\mu\gamma^\rho
	\right)\Bigr]
	\notag\\[1ex]
+\frac{1}{64} \Bigr[&
	\left(p_{h_1}^\mu-p_{h_2}^\mu\right)
	\left(g^{\nu\rho}\gamma^\sigma + g^{\nu\sigma}\gamma^\rho\right)
	\notag\\[1ex]&
	+\left(p_{h_1}^\nu-p_{h_2}^\nu\right) 
	\left(g^{\mu\rho}\gamma^\sigma +g^{\mu\sigma}\gamma^\rho\right)
	\notag\\[1ex]&
	-\left(p_{h_1}^\rho-p_{h_2}^\rho\right) 
	\left(g^{\mu\sigma}\gamma^\nu+g^{\nu\sigma}\gamma^\mu\right)
	\notag\\[1ex]&
	-\left(p_{h_1}^\sigma-p_{h_2}^\sigma\right)
	\left(g^{\mu\rho}\gamma^\nu+g^{\nu\rho}\gamma^\mu\right)
	\Bigr]
\end{align}

\begin{align}
	\label{eq:vert-hss}\notag
	&\left[\frac{\delta}{\delta\phi(p_{\phi_1})}
	\frac{\delta}{\delta\phi(p_{\phi_2})}
	\frac{\delta}{\delta h_{\mu\nu}(p_{h_1})}
	\Gamma_k\right]_{\Phi=0} \\[1ex]
	&=\frac{1}{2} \Bigr[
	g^{\mu\nu}\left(m_\phi^2 + p_{h_1}\cdot p_{\phi_1} + p_{\phi_1}\cdot p_{\phi_1}\right)
	\notag\\[1ex]&
	+p_{\phi_1}^\mu p_{h_1}^\nu + p_{\phi_1}^\nu p_{h_1}^\mu + p_{\phi_1}^\mu p_{\phi_1}^\nu
	\Bigr]
\end{align}

\begin{align}
	\label{eq:vert-hhss}\notag
	&\left[\frac{\delta}{\delta\phi(p_{\phi_1})}
	\frac{\delta}{\delta\phi(p_{\phi_2})}
	\frac{\delta}{\delta h_{ \mu \nu}(p_{h_1})}
	\frac{\delta}{\delta h_{ \rho \sigma}(p_{h_2})}	
	\Gamma_k\right]_{\Phi=0} \\[1ex]
	&=\frac{1}{4} \Bigr[
	-p_{\phi_1}^\sigma p_{\phi_2}^\mu g^{\rho\nu}
	-p_{\phi_1}^\rho p_{\phi_2}^\nu g^{\mu\sigma}
	-p_{\phi_1}^\sigma p_{\phi_2}^\nu g^{\mu\rho}
	-p_{\phi_1}^\nu p_{\phi_2}^\rho g^{\mu\sigma}
	\notag\\[1ex]&
	+p_{\phi_1}^\sigma p_{\phi_2}^\rho g^{\mu\nu}
	-p_{\phi_1}^\mu p_{\phi_2}^\sigma g^{\rho\nu}
	-p_{\phi_1}^\nu p_{\phi_2}^\sigma g^{\mu\rho}
	+p_{\phi_1}^\rho p_{\phi_2}^\sigma g^{\mu\nu}
	\notag\\[1ex]&
	+g^{\mu\sigma} g^{\rho\nu} p_{\phi_1}\cdot p_{\phi_2}
	\notag\\[1ex]&
	+g^{\rho\sigma} (-g^{\mu\nu} p_{\phi_1}\cdot p_{\phi_2}
	+p_{\phi_1}^\nu p_{\phi_2}^\mu +p_{\phi_1}^\mu p_{\phi_2}^\nu )
	\notag\\[1ex]&
	-g^{\nu\sigma} (-g^{\mu\rho} p_{\phi_1}\cdot p_{\phi_2} 
	+p_{\phi_1}^\rho p_{\phi_2}^\mu +p_{\phi_1}^\mu p_{\phi_2}^\rho)\Bigr]
\end{align}

\section{Specifying projections in the $\mathcal{X}$ sector}\label{app:proj}
The quartic tensor
structures the flow generated by the diagrams in
Fig.~\ref{fig:quartic-flow-diags} are restricted by the following requirements:
\begin{itemize}
\item All diagrams contain at least three external momenta. This is
  the case as the momentum of every external scalar must appear at the
  vertex, accounting for two of the external momenta. The third is
  required, as the chiral symmetry for the fermions requires the
  existence of a $\gamma$ matrix in the generated interaction. To
  construct a Lorentz scalar, a third momentum is then necessary to
  saturate the open index of the $\gamma$ matrix.
\item All resulting tensor structures come with momentum-dependent
  scalar legs because of the structure of the $(h)h\phi\phi$-vertex
  \eqref{eq:vert-hhss}, in which no term exists with constant scalar
  legs.  \item As the kinetic terms are reflection positive
    and Euclidean-space real operators we observe that the
    generated diagrams are reflection positive and Euclidean-space
    real as well.
        \item Further momentum structures of $\mathcal{O}(p^5)$ and
          higher are generated but will not be considered in this
          truncation as operators with more derivatives (therefore
          couplings with higher canonical dimension) are expected to
          remain less relevant at a possible fixed point.
        \end{itemize} Of the four independent reflection
          positive combinations built from the terms in
          Eq.~\eqref{eq:refPosAndEuclidImReal}, i.e.,
\begin{align}
	\mathcal{O}_{1+}&:= 
	\left(\bar\psi\gamma^\mu\nabla_\nu\psi + (\nabla_\nu\bar\psi)\gamma^\mu\psi\right)
	\left(\partial_\mu\phi\partial^\nu\phi\right)
	\\[1ex]
	\mathcal{O}_{1-}&:= 
	\left(i\bar\psi\gamma^\mu\nabla_\nu\psi - i(\nabla_\nu\bar\psi)\gamma^\mu\psi\right)
	\left(\partial_\mu\phi\partial^\nu\phi\right)
	\\[1ex]
	\mathcal{O}_{2+}&:= 
	\left(\bar\psi\gamma^\mu\nabla_\mu\psi + (\nabla_\mu\bar\psi)\gamma^\mu\psi\right)
	\left(\partial_\nu\phi\partial^\nu\phi\right)
	\\[1ex]
	\mathcal{O}_{2-}&:= 
	\left(i\bar\psi\gamma^\mu\nabla_\mu\psi - i(\nabla_\mu\bar\psi)\gamma^\mu\psi\right)
	\left(\partial_\nu\phi\partial^\nu\phi\right)
\end{align}
only the two "$-$"-combinations correspond to Euclidean-space real
operators. A complete basis of the induced flows is therefore given by
those two operators only.  Thus, while a complete basis of
$\Gamma_{k\,\,\mathcal{X}}$ contains all four independent tensor
structures, the induced action $\Gamma_{k\, \rm induced}$, 
cf. Eq.~\ref{eq:YukawaXre}, of quantum-gravity induced scalar-fermion
interactions to lowest order in a canonical power counting
is Euclidean-space real and therefore only contains
    two independent couplings.  That the flow only induces the
  $\mathcal{X}_{1/2-}$-combinations can be seen from the $\psi\bar\psi
  h$- and $\psi\bar\psi hh$-vertices in Eq.~\eqref{eq:vert-hff} and
  \eqref{eq:vert-hhff}. They both only depend on the fermionic momenta
  in the combination corresponding to the "$-$" prescription.  To show
  that this holds in all diagrams we used projection
  prescriptions on all four tensor structures and checked that
  $\dot{\mathcal{X}}_{1/2+}\rightarrow 0$ as
  $\mathcal{X}_{1/2+}\rightarrow 0$.  Then the "$+$" sector decouples
  and imaginary flows are avoided.

  To project onto the couplings
  $\mathcal{X}_{1+}$,$\mathcal{X}_{1-}$,$\mathcal{X}_{2+}$ and
  $\mathcal{X}_{2-}$, we apply functional derivatives with respect to
  the fields and switch to momentum space
  ($\partial_\mu\phi(x)\rightarrow ip_\mu\tilde\phi(p)$). For fermions
  we use the convention $\partial_\mu\psi(x)\rightarrow
  ip_\mu\tilde\psi(p)$ and $\partial_\mu\bar\psi(x)\rightarrow
  -ip_\mu\tilde{\bar{\psi}}(p)$ connected to conventions for the
  Fourier transformations fixed, e.g., in \citep{Gies:2002hq}.  To
  account for the $\gamma$-matrix we additionally project with an
  external $\gamma_\mu p_\text{ext}^\mu$ such that the Dirac trace
  does not evaluate to zero. Then we make use of the identity
  $\text{Tr}(\gamma_\mu p_a^\mu\gamma_\nu p_b^\nu)=4p_a\cdot p_b$ to
  obtain
\begin{eqnarray}
  &{}&\frac{1}{4}
  \text{Tr}\left.\left(
      \frac{\delta}{\delta\phi(p_1)}\frac{\delta}{\delta\phi(p_2)}
      \frac{\delta}{\delta\psi(p_3)}\frac{\delta}{\delta\bar\psi(p_1+p_2+p_3)}
      \Gamma_{k} \gamma_\rho
      p_\text{ext}^\rho\right)\right|_{\mathcal{O}(p^4)}
  \nonumber\\[1ex]
  &{}&=\;\mathcal{X}_{1-}\Bigl[ p_1\cdot p_\text{ext}\; p_2\cdot\left( p_1+p_2+2p_3\right) \nonumber\\[1ex]
  &{}&\,\quad\quad+\, p_2\cdot p_\text{ext}\; p_1\cdot\left( p_1+p_2+2p_3\right) \Bigr]\nonumber\\[1ex]
  &{}&-\;\mathcal{X}_{1+}\left[ p_1\cdot p_\text{ext}\; p_2\cdot\left( p_1+p_2\right) + p_2\cdot p_\text{ext}\; p_1\cdot\left( p_1+p_2\right) \right]
  \nonumber\\[1ex]
  &{}&+\;2\mathcal{X}_{2-}\left[ p_1\cdot p_2\; p_\text{ext}\cdot\left( p_1+p_2+2p_3\right) \right]\nonumber\\[1ex]
  &{}&-\;2\mathcal{X}_{2+}\left[ p_1\cdot p_2\; p_\text{ext}\cdot\left( p_1+p_2\right) \right] \quad.  
\end{eqnarray}
Projections onto $\mathcal{X}_{1-}$ and $\mathcal{X}_{2-}$ can be
determined purely by means of momentum derivatives and choice of
external momentum, i.e.,
\begin{widetext}
\begin{eqnarray}
	\mathcal{X}_{1-}&=&	\frac{-9}{16\sqrt{2}}
	\left[
          \partial_{p_1}\partial_{p_2}\partial_{p_3}\partial_{p_\text{ext}}
          \left(\text{Tr}\left[
              \frac{\delta}{\delta\phi(p_1)}\frac{\delta}{\delta\phi(p_2)}
              \frac{\delta}{\delta\psi(p_3)}\frac{\delta}{\delta\bar\psi(+p_1+p_2+p_3)}
              \Gamma_{k}\gamma_\rho p_\text{ext}^\rho
            \right]_{}\right) \right]^{p_i=0,
          \vartheta_{3,\text{ext}}=0}_{\vartheta_{1,\text{ext}}=\vartheta_{2,\text{ext}}=\sqrt{2}/3}
	\label{eq:projX3m}
	\\[1ex]
	\mathcal{X}_{2-}&=&	\frac{-9}{8\sqrt{2}}
	\left[
		\partial_{p_1}\partial_{p_2}\partial_{p_3}\partial_{p_\text{ext}}
		\left(\text{Tr}\left[
			\frac{\delta}{\delta\phi(p_1)}\frac{\delta}{\delta\phi(p_2)}
			\frac{\delta}{\delta\psi(p_3)}\frac{\delta}{\delta\bar\psi(+p_1+p_2+p_3)}	
			\Gamma_{k}\gamma_\rho p_\text{ext}^\rho
		\right]_{}\right)
	\right]^{p_i=0,\vartheta_{1,\text{ext}}=\vartheta_{2,\text{ext}}=0}_{\vartheta_{3,\text{ext}}=\sqrt{2}/3}
	\label{eq:projX4m}
	\\[1ex]
	\mathcal{X}_{1+}&=&	\frac{9i}{20}
	\left[
		\partial_{p_1}\partial_{p_2}\left(-\frac{2}{3}\partial_{p_1}+\partial_{p_2}+\partial_{p_3}\right)\partial_{p_\text{ext}}
		\left(\text{Tr}\left[
			\frac{\delta}{\delta\phi(p_1)}\frac{\delta}{\delta\phi(p_2)}
			\frac{\delta}{\delta\psi(p_3)}\frac{\delta}{\delta\bar\psi(+p_1+p_2+p_3)}	
			\Gamma_{k}\gamma_\rho p_\text{ext}^\rho
		\right]_{}\right)
	\right]^{p_i=0,\vartheta_{3,\text{ext}}=0}_{\vartheta_{1,\text{ext}}=\frac{1}{2},\vartheta_{2,\text{ext}}=\frac{1}{3}}
	\label{eq:projX3p}
	\;.
\end{eqnarray}
\end{widetext}
For concreteness we give an explicit parametrization of such a
momentum configuration
\begin{align}
	\label{eq:fullySymm}
	p_1 = \left(\begin{array}{c}
		1\\
		0\\
		0\\
		0\\
	\end{array}\right)
	,\;
	p_2 = \left(\begin{array}{c}
		-1/3\\
		2\sqrt{2}/3\\
		0\\
		0\\
	\end{array}\right)
	,\;
	p_3 = \left(\begin{array}{c}
		-1/3\\
		-\sqrt{2}/3\\
		-\sqrt{2/3}\\
		0\\
	\end{array}\right)\,.
\end{align}

The explicit external momentum choices corresponding to
the above explicit parametrization (cf.~\eqref{eq:fullySymm}) and
the projections \eqref{eq:projX3m}~-~\eqref{eq:projX3p} are
\begin{eqnarray}
	p_{\text{ext},\mathcal{X}_{1-}} &=&
	\left(\begin{array}{c}
		\sqrt{2}/3\\
		2/3\\
		-1/\sqrt{3}\\
		0
	\end{array}\right)
	,\;
	p_{\text{ext},\mathcal{X}_{2-}} = 
	\left(\begin{array}{c}
		0\\
		0\\
		-1/\sqrt{3}\\
		\sqrt{2/3}
	\end{array}\right)
	,\;\notag\\
	p_{\text{ext},\mathcal{X}_{1+}} &=&
	\left(\begin{array}{c}
		1/2\\
		3/(4 \sqrt{2})\\
		-5/(4 \sqrt{6})\\
		-\sqrt{5/24}
	\end{array}\right)
	\,.
\end{eqnarray}

Assuming the same fully symmetric momentum configuration and allowing
for the most general derivative projection, i.e., $(A\partial_{p_1} +
B\partial_{p_2}
+ \partial_{p_3})\partial_{p_1}\partial_{p_2}\partial_{p_\text{ext}}$,
the generated system of equations does not allow for a solution
defining a clean $\mathcal{X}_{2+}$-projection.  Therefore we resort
to a $\mathcal{X}_{2+}$-projection corresponding to a non-symmetric
momentum configuration. The according momenta read
\begin{align}
	\label{eq:fullySymm1}
	p_1^\mu =& p_1\left(\begin{array}{c}
		1\\
		0\\
		0\\
		0\\
	\end{array}\right)
	,\;
	p_2^\mu = p_2\left(\begin{array}{c}
		-1/3\\
		2\sqrt{2}/3\\
		0\\
		0\\
	\end{array}\right)
	,\notag\\
	p_3^\mu =& p_3\left(\begin{array}{c}
		-1/3\\
		-7/\sqrt{72}\\
		\sqrt{5/24}\\
		0\\
	\end{array}\right)
	,\;
	p^\mu_{\text{ext},\mathcal{X}_{2+}} = p_{\text{ext}}
	\left(\begin{array}{c}
		-1/3\\
		5/\sqrt{72}\\
		\sqrt{13/24})\\
		0
	\end{array}\right)
	\,.
\end{align}
with a projection prescription
\begin{widetext}
\begin{align}
\mathcal{X}_{2+}=	i
	\left[
		\partial_{p_1}\partial_{p_2}\left(\frac{5}{16}\partial_{p_1}+\frac{7}{16}\partial_{p_2}-\frac{9}{2(-9+\sqrt{65})}\partial_{p_3}\right)\partial_{p_\text{ext}}
		\left(\text{Tr}\left[
			\frac{\delta}{\delta\phi(p_1)}\frac{\delta}{\delta\phi(p_2)}
			\frac{\delta}{\delta\psi(p_3)}\frac{\delta}{\delta\bar\psi(+p_1+p_2+p_3)}	
			\Gamma_{k}\gamma_\rho p_\text{ext}^\rho
		\right]_{}\right)
	\right]
	\label{eq:projX4p}
	\;.
\end{align}
\end{widetext}

\section{Analytic $\beta$-functions in the $\mathcal{X}$-sector}
\label{app:Xbetas}

Here we present the full $\beta$-functions in the
$\mathcal{X}$-sector, as obtained with the above projections,
cf. App.~\ref{app:proj}, evaluated at $\mathcal{X}_{1/2+}\rightarrow
0$.

The two ``$+$''-combinations are not induced by the flow and thus
remain as trivial couplings. Independent of metric fluctuations their
Gau\ss ian fixed points persist and they pose a trivial extension to the
solutions discussed in Sec.~\ref{sec:fully_disentangled_truncation}.

\begin{widetext}

\begin{align}
\dot{\mathcal{X}}_{1-}\Big{|}_{\mathcal{X}_{1/2+}\rightarrow 0} =
&
 (4+\eta_{\psi,0}+\eta_{\phi,0})\mathcal{X}_{1-} 
-\frac{2 (\eta _h-7) g_n{}^2}{7 (\mu _h+1){}^3}-\frac{10 (\eta _h-6) g_n{}^2}{9 (\mu _h+1){}^3}-\frac{g_n{}^2 (\eta _{\psi }-6)}{9 (\mu _h+1){}^2}-\frac{5 (\eta _h-8) g_n \mathcal{X}_{1-}}{384
   \pi  (\mu _h+1){}^2}+\frac{g_n (\eta _{\psi }-7) \mathcal{X}_{1-}}{84 \pi  (\mu _h+1)}
\notag\\[1ex]&
+\frac{(\eta _h-7) g_n (11 \mathcal{X}_{1-}-52 \mathcal{X}_{2-})}{840 \pi  (\mu _h+1){}^2}+\frac{(\eta
   _h-6) g_n (17 \mathcal{X}_{1-}+8 \mathcal{X}_{2-})}{96 \pi  (\mu _h+1){}^2}+\frac{g_n (\eta _{\psi }-6) (9 \mathcal{X}_{1-}-8 \mathcal{X}_{2-})}{360 \pi  (\mu _h+1)}
\notag\\[1ex]&
+\frac{(\eta _{\psi }-8) \left(-5 \mathcal{X}_{1-}{}^2+8 \mathcal{X}_{1-}
   \mathcal{X}_{2-}+4 \mathcal{X}_{2-}{}^2\right)}{1792 \pi ^2}
-\frac{(\eta _{\phi }-9) \left(121 \mathcal{X}_{1-}{}^2+64 \mathcal{X}_{1-} \mathcal{X}_{2-}+4 \mathcal{X}_{2-}{}^2\right)}{6048 \pi
   ^2}
\label{eq:betaX3m}
\\[2ex]
\dot{\mathcal{X}}_{2-}\Big{|}_{\mathcal{X}_{1/2+}\rightarrow 0} =
&
 (4+\eta_{\psi,0}+\eta_{\phi,0})\mathcal{X}_{2-} 
+\frac{(\eta _h-7) g_n{}^2}{14 (\mu _h+1){}^3}+\frac{5 (\eta _h-6) g_n{}^2}{18 (\mu _h+1){}^3}+\frac{g_n{}^2 (\eta _{\psi }-6)}{36 (\mu _h+1){}^2}
\notag\\[1ex]&
-\frac{(\eta _h-8) g_n (7 \mathcal{X}_{1-}+33
   \mathcal{X}_{2-})}{384 \pi  (\mu _h+1){}^2}-\frac{(\eta _h-7) g_n (2 \mathcal{X}_{1-}-7 \mathcal{X}_{2-})}{120 \pi  (\mu _h+1){}^2}+\frac{(\eta _h-6) g_n (4 \mathcal{X}_{1-}+13
   \mathcal{X}_{2-})}{96 \pi  (\mu _h+1){}^2}
\notag\\[1ex]&
-\frac{g_n (\eta _{\psi }-7) (7 \mathcal{X}_{1-}+24 \mathcal{X}_{2-})}{336 \pi  (\mu _h+1)}-\frac{g_n (\eta _{\psi }-6) (9 \mathcal{X}_{1-}-44
   \mathcal{X}_{2-})}{2880 \pi  (\mu _h+1)}+\frac{g_n (\eta _{\psi }-6) (44 \mathcal{X}_{2-}-9 \mathcal{X}_{1-})}{2880 \pi  (\mu _h+1)}
\notag\\[1ex]&
+\frac{(\eta _{\phi }-9) \left(59 \mathcal{X}_{1-}{}^2-52
   \mathcal{X}_{1-} \mathcal{X}_{2-}-76 \mathcal{X}_{2-}{}^2\right)}{12096 \pi ^2}+\frac{(\eta _{\psi }-8) (\mathcal{X}_{1-}-2 \mathcal{X}_{2-}) (\mathcal{X}_{1-}+\mathcal{X}_{2-})}{896 \pi ^2}
   \label{eq:betaX4m}
\\[2ex]
\dot{\mathcal{X}}_{1+}\Big{|}_{\mathcal{X}_{1/2+}\rightarrow 0} = &0
\\[1ex]
\dot{\mathcal{X}}_{2+}\Big{|}_{\mathcal{X}_{1/2+}\rightarrow 0} = &0
\end{align}

\end{widetext}

The vanishing flows $\dot{\mathcal{X}}_{1/2+} \equiv 0$ guarantee that the ``$+$''-sector fully decouples and remains zero at throughout the flow.

\bibliography{../YukawaGravity}

\end{document}